\documentclass{aa}

\usepackage{graphicx}
\usepackage{txfonts}
\usepackage{amsmath}
\usepackage{placeins}
\usepackage{xcolor}
\usepackage{longtable}

\usepackage[colorlinks=true,linkcolor=blue,citecolor=blue,urlcolor=blue]{hyperref}

\usepackage{lipsum}
\usepackage{subcaption}
\usepackage{lscape}
\usepackage{lineno}

\title{The Large Array Survey Telescope---Pipeline.}
\subtitle{II. Image Subtraction and Transient Detection}

\titlerunning{LAST Pipeline~II}
\authorrunning{R. Konno et al.}

\author{
R.~Konno\inst{1}
\and E.~O.~Ofek\inst{1}
\and A.~Krassilchtchikov\inst{1}
\and Y.~Shvartzvald\inst{1}
\and S.~Ben-Ami\inst{1}
\and D.~Polishook\inst{1}
\and C.~Tishler\inst{1}
\and E.~Segre\inst{1}
\and S.~Garrappa\inst{1}
\and E.~A.~Zimmermann\inst{1}
\and A.~Horowicz\inst{1}
\and P.~Chen\inst{1,2}
\and A.~Gal-Yam\inst{1}
\and M.~Engel\inst{1}
\and Y.~M.~Shani\inst{1}
\and S.~A.~Spitzer\inst{1}
\and S.~Fainer\inst{1}
\and O.~Yaron\inst{1}
\and A.~Blumenzweig\inst{1}
}

\institute{
Department of Particle Physics and Astrophysics, Weizmann Institute of Science, 76100 Rehovot, Israel
\and 
Institute for Advanced Study in Physics, Zhejiang University, Hangzhou 310027, China
}

\begin{document}

\abstract
{The Large Array Survey Telescope (LAST) is a wide-field visual-band survey designed to explore the variable and transient sky with high cadence. Its raw data stream is automatically processed in near real time at the observatory site, producing science-quality images, catalogs, and transient alerts. Transient alerts are then reported to the Transient Name Server (TNS).}
{The LAST pipeline comprises two major components: (i) processing and calibration of single images followed by coaddition of $20\times20$\,s exposures, producing single-image and coadded-image catalogs; and (ii) subtraction of coadded images from calibrated reference images followed by transient detection. In this work we present a detailed description and validation of the second component of the pipeline.}
{Transient detection is based on the algorithm for proper image subtraction (ZOGY). We combine ZOGY subtraction with the \textsc{Translient} statistic for sub-pixel motion discrimination, together with a sequence of deterministic filtering steps, to produce a clean stream of transient candidates without the use of machine learning.}
{Using commissioning data, the pipeline achieves a preliminary $5\sigma$ limiting magnitude of $20.3$–$20.7$\,mag, a single-epoch transient detection efficiency of $\sim80\%$, and a purity of $\gtrsim90\%$ at signal-to-noise ratio of $\geq7.5\sigma$.}{}

\keywords{
methods: data analysis ---
methods: observational ---
techniques: image processing ---
techniques: photometric ---
telescopes ---
transients}
\maketitle 

\section{Introduction}
\label{sec:Introduction}

Sky surveys have become a major enabling force for astrophysical discoveries over the last two decades (e.g., \citealt{Quimby+2011_SLSN,  Drout+2014_Rapidly_Evolving, Gal-Yam+2014_SN2013cu_FlashSpectroscopy, Ho+2023Natur_MinuteTimeScaleFlares_Transient}). However, with the large amount of data acquired by such surveys, efficient data processing has become necessary for prompt achievement of scientific goals.

The Large Array Survey Telescope (LAST) is a new cost-effective visible-band instrument (\citealt{Ofek+BenAmi2020_Grasp_SkySurvrys_CostEffectivness}) currently being commissioned at the Weizmann Astrophysical Observatory (WAO\footnote{\url{https://www.weizmann.ac.il/wao}}) in the Arava desert of Southern Israel. The overall LAST system design is discussed in \cite{Ofek+2023PASP_LAST_Overview}, while its science goals are outlined in \cite{BenAmi+2023PASP_LAST_Science}. Once finalized, LAST will consist of $72$ $28$-cm f/$2.2$ telescopes. A total of 40 telescopes are already operational. Each LAST telescope carries a $\sim61$\,Mpix ($6400\times9600$ pixels) backside-illuminated CMOS camera with $3.6\,\mu$m pixels. The telescope and camera combination produces a pixel scale of $1.25''\,$pix$^{-1}$ and a field of view (FoV) of $\cong 7.4\,$deg$^{2}$ per telescope. Four telescopes are attached to a single mount, totaling $18$ units in the $72$-telescopes configuration. This makes LAST a flexible sky survey that can simultaneously cover 530\,deg$^{2}$, or alternatively, use multiple telescopes to observe the same pointing position. In the fully co-aligned configuration, the effective collecting area of LAST is equivalent to a 2.4\,m ($\cong 0.28\times\sqrt{72}$) telescope with a 7.4\,deg$^{2}$ FoV. 

Data processing in LAST is organized in two parts. The first part of the LAST pipeline (Pipeline~I), which performs source detection and provides photometric and astrometric measurements of the detected sources, is described in \cite{Ofek+2023PASP_LAST_PipeplineI}. The second part (Pipeline~II) operates on the calibrated products from Pipeline~I to perform image subtraction and detection of transient astrophysical phenomena, hereafter transients. Here we describe and assess the performance of Pipeline~II. The composite pipeline is written in MATLAB and is distributed as part of the AstroPack package\footnote{\url{https://github.com/EranOfek/AstroPack}}.

Image subtraction and identification of transients is carried out by many surveys. Some prominent examples are the Zwicky Transient Facility (\citealt{Bellm+2019_ZTF_Overview, Graham+2019_ZTF_objectives, Masci+2019_ZTF_Pipeline}), Pan-STARRS (\citealt{Chambers+2016_PS1_Surveys}), ATLAS (\citealt{Tonry2011_ATLAS_SurveyCapability, Heinze+2018_ATLAS_VarStars}),
GOTO (\citealt{Steeghs+2022MNRAS_GOTO_TelescopeSurvey}), ASAS-SN  \citep{Kochanek+2017_ASASSN_VarStars}, and the Vera C. Rubin Observatory Legacy Survey of Space and Time (LSST; \citealt{Ivezic+2019_LSST_Survey}). Historically, image subtraction over a large FoV has resulted in a large number of false alarms.
The reasons for these false alarms are diverse, but generally due to one of the following categories: numerically unstable operations; incorrect error propagation; physical artifacts like bad pixels, cosmic rays, and ghosts; and employment of formalism based on inaccurate assumptions, e.g. that the images are perfectly registered or flux matched. 
Finding real transients among the large number of false alarms is an important problem that has received significant attention and effort by means of software, hardware, and human power within the last two decades. In the early days of the Palomar Transient Factory survey (\citealt{Law+2009_PTF}), this task was performed by human scanners that sifted through only the best hundreds or thousands of transient candidates per day. These days, the method of choice in most surveys is to employ machine learning (ML) tools to separate between real and bogus events (e.g., \citealt{Mahabal+2019PASP_MachineLearning_ZTF,Forster+2021AJ_ALeRCE_MachineLearning_Transients,vanRoestel+2021AJ_MachineLearning_TransientsClass_ZTF}). ML methods efficiently classify complex residuals but are often opaque and require extensive retraining as survey conditions evolve. In contrast, deterministic approaches exploit physically motivated features available across the full dataset, beyond local image stamps, and remain interpretable and stable under changing conditions. Therefore, deterministic transient detection methods are still required at least as an additional step to ML or even as a viable alternative.

For the LAST Pipeline~II, the main approaches employed to transient detection and vetting include the proper image subtraction (ZOGY) (\citealt{Zackay+2016_ZOGY_ImageSubtraction}), the {\sc Translient} sub-pixel motion discriminator (\citealt{Springer+Ofek+2024AJ_Translient}), and several additional tools. Currently, no ML component is used. Our approach avoids black-box classification and instead targets the physical origins of false positives using deterministic features derived from the data.

In order to evaluate and compare the capabilities of various approaches to transient detection, a uniform and statistically motivated method should be used. We suggest that such a metric is a combination of efficiency and purity as a function of the signal-to-noise ratio ($S/N$). Unfortunately, for most sky surveys, purity and efficiency are not reported as a function of $S/N$. In this work, injections of artificial transients into real data are used to measure the transients retention efficiency. The purity is instead estimated from real data directly.

We start by analyzing major sources of image noise that produce subtraction residuals and then construct statistical and heuristic methods to identify and treat such residuals. We then estimate the efficiency of our method as a function of $S/N$ by using simulated transients, while the purity is derived as an integral over $S/N$ by verifying each passing real transient candidate. We also derive the expectation values for the yearly transient rates and compare against expected false alarm rates. While the pipeline will continue to evolve, the core statistical framework and feature space described here are expected to remain stable and form the basis for future developments.

In \S\ref{sec:Overview} we provide a brief overview of the major data reduction
stages of the first and the second parts of the LAST pipeline. In \S\ref{sec:SubtractionProducts} and \S\ref{sec:FalsePositives} we provide a detailed description of the second part of the pipeline. In \S\ref{sec:performance} we assess the performance of the pipeline. In \S\ref{sec:conclusion} we formulate conclusions and provide an outlook for future development of the pipeline.

\section{Pipeline Overview}
\label{sec:Overview}
The LAST data reduction process (Pipeline~I) is described in detail in \cite{Ofek+2023PASP_LAST_PipeplineI}. A step-by-step summary is provided in Appendix~\ref{sec:appendix_pipeI}. In short, Pipeline~I partitions a full-frame LAST image into $24$ sub-images and performs standard image calibration, source detection, and photometric and astrometric measurements. The sub-images have a size of $1726\times1726$ pixels, including a 64-pixel-wide overlap between sub-images.

The sub-images are coadded in $20\times20\,$s exposures and further coaddition products are derived. These products consist of the calibrated $20\times20~$s sub-images, $N$, their associated source catalogs, bit-mask images, and point-spread function (PSF) model stamps, $P_N$. The bit-mask images carry bit-coded diagnostics concerning each pixel, such as whether a pixel is saturated or is within the overlap region.

\begin{figure*}
	\centering
	\includegraphics[scale=0.60]{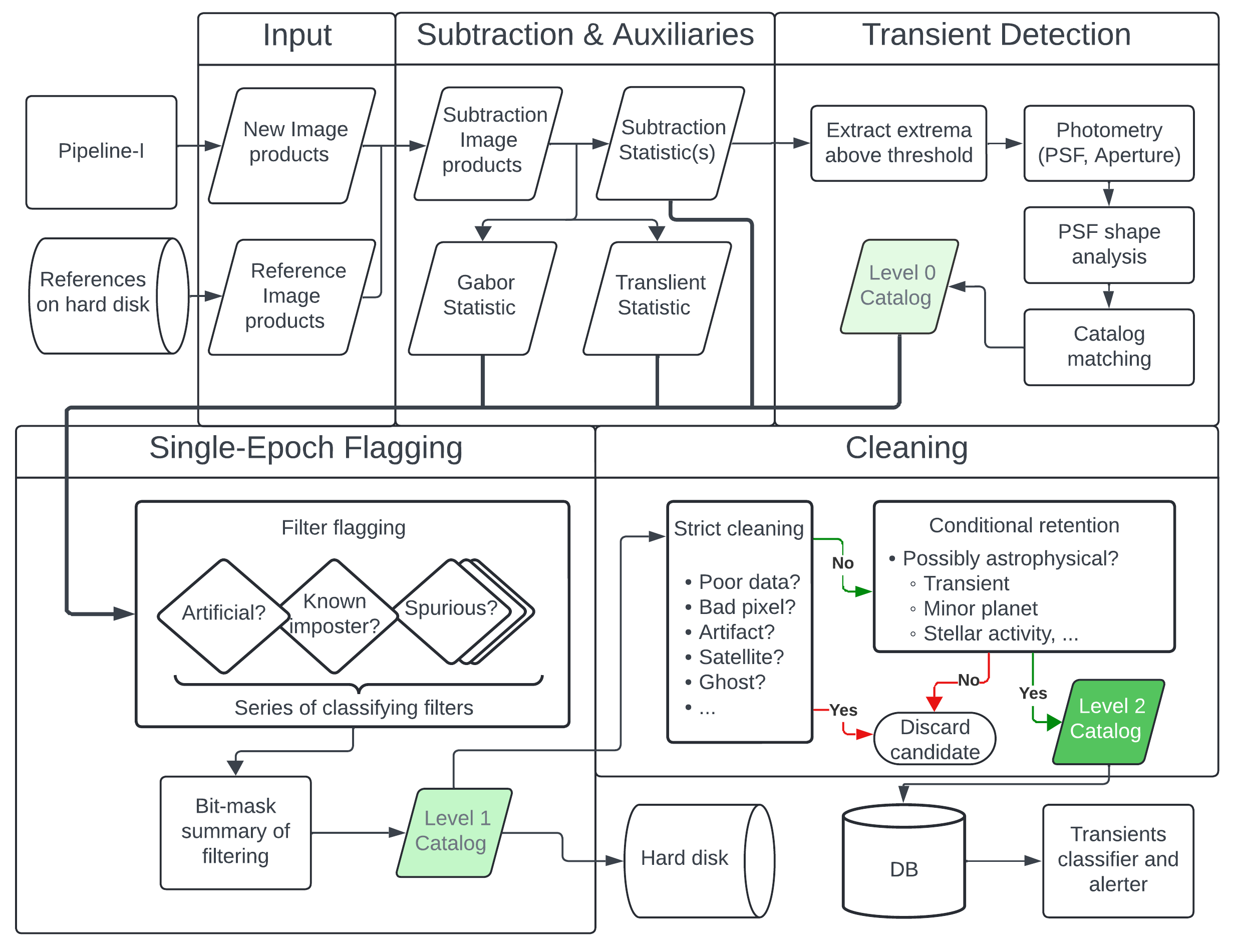} 
	\caption{Flowchart of Pipeline~II responsible for LAST transients detection and filtering.}
	\label{fig:LAST_transients_pipe}
\end{figure*}

The coadd products are ingested by Pipeline~II, which then executes the image subtraction and transients detection process. A flowchart of Pipeline~II is illustrated in Figure~\ref{fig:LAST_transients_pipe}. The primary steps of Pipeline~II are as follows:

\begin{enumerate}
    \item Find and load a pre-constructed deep reference image $R$ into memory. The $R$ image is accompanied by a bit-mask image, a PSF stamp $P_R$, and a source catalog.
    \item Geometrically register the reference image $R$ and new calibrated image $N$ to the same frame. Currently, this is done using cubic interpolation.
    \item Perform image subtraction between the $R$ and $N$ images via the ZOGY method \citep{Zackay+2016_ZOGY_ImageSubtraction} to produce a difference image $D$ and its PSF $P_D$. 
    \item Propagate the $R$ and $N$ bit-mask images to the $D$ bit-mask image via the {\it or} operator.
    \item Match-filter the $D$ image with its PSF to produce the ZOGY $S$ statistics image.
    \item Match-filter the $D$ image with a $\delta$-function to produce the $S_\delta$ statistics image.
    \item Match-filter the $D$ image with an extended PSF to produce the $S_\mathrm{Ext}$ statistics image.
    \item Calculate the \textsc{Translient} statistics image (\citealt{Springer+Ofek+2024AJ_Translient}).
    \item Match-filter the $D$ image with the {\it Gabor} filter \citep{Gabor1946TheoryOC} to produce the $S_\text{g}$ statistic image.
    \item Extract all positive and negative sources with $\vert S \vert>5\sigma$. These sources constitute the initial candidates list.
    \item Perform PSF-fitting photometry on all initial candidates within the $R$, $N$, and $D$ images.
    \item Measure the shapes of $P_N$ and $P_R$, as well as the local PSF shapes of each candidate in $D$. 
    \item Perform aperture photometry on all initial candidates within the $D$ image. The apertures used have radii of $2$, $4$, and $6$ pixels.
    \item Cross-match the single-epoch candidates list with known Solar System objects.
    \item Cross-match the single-epoch candidates list with external catalogs.
    \item Apply filters on the initial candidates in order to reduce the number of false positives.
    \item Store the full list of single-epoch candidates to disk.
    \item Remove from the list all candidates associated with poor data and severe artifacts and inject catalog data on the remaining objects into the transients database.
\end{enumerate}

The injected catalog of single-epoch candidates is the final data product of Pipeline~II. Due to the overlap between neighboring sub-images, the same transient candidate may be detected multiple times. In such cases, duplicate detections are resolved by retaining the candidate whose position is closest to the center of its respective sub-image, thereby minimizing edge-related effects. Further treatment of the injected transient candidates, such as classification and reporting, is carried out by external processes. Currently, the sub-selected candidates injected into the database are queried against the same database in order to retrieve multi-epoch detection history. All candidates which pass either a single-epoch significance threshold or show multiple detections are reported internally. A more extensive transient-classification pipeline is under development and will be described elsewhere.

The sequence of steps outlined above defines the core structure of Pipeline~II and is expected to remain stable, while individual components, e.g., filtering thresholds or classification layers, may be further optimized.

\section{Generation of subtraction products}
\label{sec:SubtractionProducts}

In this section and the next, we provide a detailed description of selected steps of Pipeline~II described in \S\ref{sec:Overview}. First, we describe the construction of the subtraction products, including the registration, subtraction, and measurement steps.

\subsection{Reference images}

LAST partitions the whole sky into 1784 fields. This partition corresponds to the FoV of a single four-telescope mount unit with an overlap of a few arcminutes between neighboring fields. The reference images are currently constructed per field. To keep the images small enough for efficient processing, the reference images correspond to the sub-image size. This means that for each field, each of the four telescopes has 24 sub-image reference images, i.e., there are 96 reference images per LAST field.

The reference images are currently generated using a weighted coaddition scheme. In the future, we plan to employ the proper coaddition method (\citealt{Zackay+2017_CoadditionI, Zackay+2017_CoadditionII}). Additionally, we are revising the reference image framework to provide continuous sky coverage independent of the current field partitioning. In this scheme, reference images will be generated such that any $N$ image, regardless of its exact pointing or orientation, is fully covered by a valid reference region.

\subsection{Geometric Registration}
\label{sec:Registration}

For a new calibrated science image $N$, the corresponding reference image $R$ is loaded. $R$ is then registered to $N$ geometrically. Here, the new $R$ pixel grid is derived by converting the pixel grid of $N$ into sky coordinates via the world coordinate system (WCS) and then converting the sky coordinates into pixel coordinates of $R$. The count numbers within the new grid pixels are then derived by resampling the previous $R$ pixel grid with a cubic interpolation. The mask image is interpolated with a nearest-neighbor interpolation.
Extrapolated values outside of the original $R$ pixel grid are set to \texttt{NaN}.
In addition, the pixels outside the $R$ grid are marked in the $R$ bit-mask image with the \texttt{NaN} flag.

Since LAST observations, as well as the $R$ images, are segmented into pre-defined fields, and the telescope pointing is not perfect, deviations occur between each pointing, measured at a median of $16\,$pix ($20.0''$) in image $x$ and $y$ coordinates. These deviations lead to trimming of the $R$ image during registration. In effect, two sides of the $R$ image are truncated, 
while the opposite sides have \texttt{NaN}-valued borders. To minimize the change in the histogram and frequency spectrum of the $R$ image, the \texttt{NaN}-valued borders are replaced by sampling a normal distribution defined by the mean and the standard deviation of the $R$ background rate.
However, in the $R$ bit-mask image, these pixels are still marked as \texttt{NaN}. 

These edge effects arise primarily from the finite spatial extent and fixed tiling of the reference images, combined with pointing variations, and are not fully mitigated by the overlap between sub-images. A sky-grid independent reference scheme, currently in development, will largely eliminate such edge effects.

\subsection{Flux normalization}

The integrated flux between two images of the same field varies with observing conditions and must be placed on a consistent relative photometric scale for accurate image subtraction. Generally, this may be done by an iterative flux matching process, which increases computational costs. A more efficient method, which we also adopt, is to handle the relative scaling via the photometric zero-point calibration of each image. The zero points of $N$ and $R$ are used as explicit arguments in the subtraction method. Uncertainties in the flux normalization are further discussed in \S\ref{sec:flux_ratio_unc}.

\subsection{Image subtraction}

Image differencing is performed with the ZOGY algorithm (\citealt{Zackay+2016_ZOGY_ImageSubtraction}). Its main products are: (i) a proper subtraction image $D$, which is characterized by independent and identically distributed (i.i.d.) noise\footnote{In fact, the noise is not i.i.d. on top of bright sources, as source noise is correlated.}; (ii) a $D$-image PSF $P_{\rm D}$; and (iii) a statistic score image $S$, obtained by match-filtering (cross-correlating) $D$ with $P_{\rm D}$. The $S$ statistic tests for PSF-like residuals in $D$ and therefore rises on the position of variable and transient sources. 

However, $S$ does not account for astrometric or source noise, so moving or bright persistent sources will also produce a positive signal. \cite{Zackay+2016_ZOGY_ImageSubtraction} proposed estimating these noise terms and penalizing $S$ to form the corrected statistic $S_\text{corr}$. Accordingly, our subtraction step also produces an $S_\text{corr}$ image where the source noise is accounted for. Astrometric noise is instead treated by the \textsc{Translient} algorithm (see \S\ref{sec:Translient}) and is not considered by $S_{\rm corr}$. 

We additionally match-filter the $D$ image with a $\delta$-function kernel and produce a $\delta$-function score image $S_{\delta}$. This is used later in the filtering process for cosmic rays and hot pixels identification, as in these cases we expect $S_{\delta}>S$. Hot pixels may survive the coaddition process as the images of a visit are taken in a continuous video mode with no dead time and are thus not dithered.

Similarly, we convolve $P_{\rm D}$ with a Gaussian kernel and repeat the match-filter process to construct an extended-PSF score image $S_{\rm Ext}$. In this case, the comparison with $S$ allows the filtering of extended objects such as comets. All statistics images are normalized by the standard deviation and given in $S/N$ units of $\sigma$.

\subsection{Initial candidates selection}
Given a proper subtraction statistic image $S$, a search for local extrema across the image is performed. All the subtraction residuals with $\vert S\vert \geq 5\sigma$ are considered as transient candidates, either rising or fading. The initial list of unfiltered transient candidates is referred to as the level $0$ catalog.

\subsection{Photometry}

For each initial level $0$ transient candidate, we perform PSF fitting in images $N$, $R$, and $D$ using their respective PSFs, $P_{\rm N}$, $P_{\rm R}$, and $P_{\rm D}$. Since $D$ is an image with an i.i.d. noise, the fitting problem is diagonalized and there is no need for a covariance matrix. In the fitting process, there are three free parameters: the image coordinates $x$ and $y$ as well as the flux. Generally, we use a small fitting radius of 2 or 3 pixels. Fitting radius means that all pixels found within this radius from the source center are participating in the chi-square goodness-of-fit $\chi^{2}$ calculation. The fitter includes a position-drift-limiter that prevents the best fit position to drift by more than 1 pix, compared to the initial position. For each fit, this process results in the flux, the flux error, the PSF-fit $(S/N)_{\rm PSF}$, and $\chi^{2}$. 

Additionally, aperture photometry is applied on each candidate position in the $D$ image. Three different apertures of pixel radii $2$, $4$, and $6$ are used. The local background is estimated from a surrounding annulus whose inner and outer radii are set dynamically according to the available PSF stamp size, with minimum bounds imposed to keep the annulus outside the core and sufficiently wide.

\section{False-positive sources and their treatment}
\label{sec:FalsePositives}
We now turn to the filtering stage. We describe the dominant sources of false positive transient candidates and the methods used to distinguish them from real transients.

\subsection{Translient}
\label{sec:Translient}

In general, image subtraction methods are based on various assumptions. Violations of these assumptions lead to artificial subtraction residuals. One such assumption is that the images are perfectly registered. This assumption is always violated. For example, in ground-based images, due to the atmospheric astrometric scintillation (e.g., \citealt{Lindegren1980_AtmosphericScintilation_GroundBasedAstrometry, Shao+Colavita1992_AtmosphericScintilation_GroundBasedAstrometry, Ofek2019_Astrometry_Code}) the stars' positions jitter around their mean apparent position with amplitude larger than the Poisson noise jitter. This means that the registration to the accuracy level of the Poisson noise is not possible.
\cite{Springer+Ofek+2024AJ_Translient} suggested an optimal\footnote{\textsc{Translient} is an optimal statistic when the shifts are $\ll$\,pixel.} solution to this problem named \textsc{Translient}. In simple terms, the \textsc{Translient} statistic is designed to identify subtraction residuals that are caused by small positional shifts between the $N$ and $R$ images, e.g., due to imperfect registration or atmospheric jitter.

By comparing the \textsc{Translient} statistics to the ZOGY statistics for each candidate we estimate whether the residual is better explained by apparent motion or by intrinsic variability. Since the ZOGY and \textsc{Translient} hypotheses are not nested and have different numbers of degrees of freedom, the comparison is performed using the Akaike information criterion (AIC).
Since AIC is only asymptotically correct for large sample sizes,
we are using the corrected formula (AIC$_\text{c}$) for small sample size and Gaussian i.i.d. noise \citep{AICc1993}:
\begin{equation}
    \text{AIC}_{\rm c} = 2k - 2\ln{L} + \frac{2k^2+2k}{n-k-1}.
\end{equation}
Here $n$ is the sample size ($=1$), $k$ is the number of degrees of freedom, and $L$ is the test statistic. The \textsc{Translient} model has two internal degrees of freedom, while the ZOGY model has one.

\subsection{The Gabor filter}

Since the $D$ image is produced in the Fourier space, it is susceptible to Gibbs ringing when transformed back into the real space. This mainly affects the vicinity of saturated stars where a step function in the image data is present. I.e., in saturated regions, the image is not Nyquist sampled. To identify the regions affected by Gibbs ringing, the $D$ image is match-filtered with a Gabor filter \citep{Gabor1946TheoryOC}. A Gabor filter is a Gaussian envelope modulated by a sinusoid, and it has the form:
\begin{eqnarray}
    g(x,y) &=& \frac{1}{2\pi\sigma_x\sigma_y}e^{-\frac{1}{2}(\frac{x^2}{\sigma_x^2}+\frac{y^2}{\sigma_y^2})}e^{-2\pi i((u_0x+v_0y)+\phi)},
\end{eqnarray}

\noindent where ($x,y$) are the pixel coordinates, ($\sigma_x,\sigma_y$) are the Gaussian standard deviations in the x and y directions, ($u_0,v_0$) are the sinusoidal frequencies in the x and y directions, and $\phi$ is the sinusoidal phase. For the sake of efficiency, a single filter is used corresponding to a symmetric ringing artifact on the shortest wavelengths, i.e., $\sigma_x = \sigma_y = 1$, $u_0=v_0=1/2$, and $\phi=0$. This process produces a Gabor-filter statistics image $S_\text{g}$. In practice, $S_\text{g}$ is additionally normalized by its standard deviation.

\subsection{Flux ratio uncertainty}
\label{sec:flux_ratio_unc}
In general, it is difficult to photometrically calibrate ground-based astronomical images to better than $\sim1.5\%$ accuracy \citep{Padmanabhan+2008_SDSS_ImprovedPhotometricCalibration, Ofek+2012_photCalib, Ofek+2012_PTF_photCat, Schlafly+2012_PS1_PhotometricCalibration, Masci+2019_ZTF_Pipeline}. However, \cite{SimonePhotometryCalib} demonstrated that fitting the photometric transmission for each image can achieve a typical accuracy of $3$–$8\,$mmag, though this method is not yet implemented in the LAST pipeline. Even so, at $\sim1\%$ photometric accuracy, the noise for sources with flux below $10\,000\,$e$^{-}$ remains dominated by Poisson statistics. The calibration uncertainty of the zero points then propagates into the subtraction process.

Additionally, the photometric zero point depends on the color and airmass of the calibration sources. \cite{SimonePhotometryCalib} measured this dependence for LAST at $\sim 0.01-0.05$\,mag\,mag$^{-1}$\,airmass$^{-1}$, where mag$^{-1}$ denotes color given in magnitudes. As a result, the calibrated zero point is biased toward the average color of the calibration sample. As airmass varies between exposures, this bias is not constant from image to image and introduces an error in the flux normalization ratio for persistent sources whose colors deviate from the calibration samples. This yields larger subtraction residuals, particularly on top of stars. These effects are particularly pronounced for LAST, which operates without a filter and thus has a broad effective bandpass, enhancing color-dependent atmospheric and instrumental systematics. The discussion and the solution to this problem will be presented in a dedicated publication (Konno et al., in prep.).

\subsection{PSF reconstruction errors}
\label{sec:PSF_reco_errors}

\begin{figure}
\centerline{\includegraphics[width=0.45\textwidth]{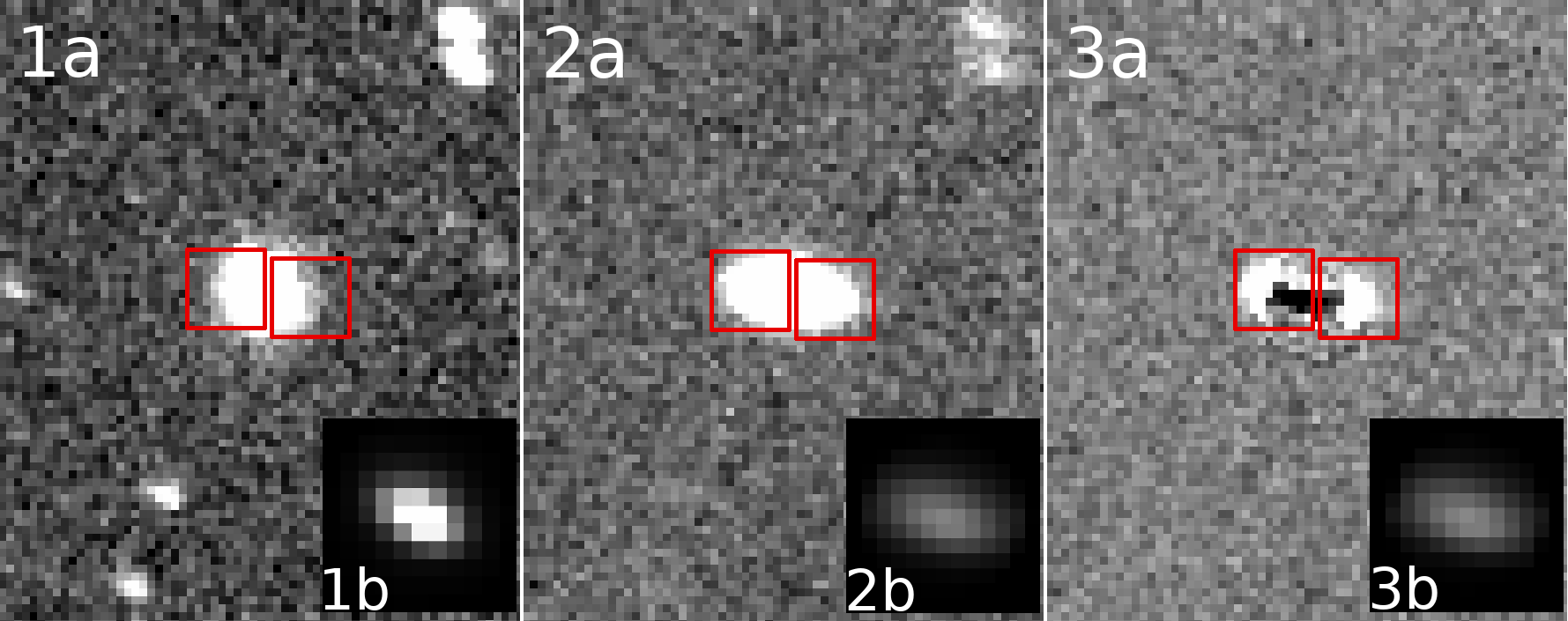}}
    \caption{Example of an artifact caused by an elongated PSF. Shown are reference $R$, new $N$, and difference $D$ images labeled as 1--3a, respectively. Also shown are their corresponding PSFs labeled as 1--3b. The PSF color scale is fixed to be identical. In this case, image $N$ was taken during strong wind conditions ($>20\,$km$\,$h$^{-1}$), which led to  shaking of the mount and elongation of the PSF. The subtraction then results in an excess at both ends of the elongated PSF axis and to a false detection of two transient candidates (red squares).}
\label{fig:WindSubAndPSF}
\end{figure}

The ZOGY subtraction method is robust to arbitrary PSF shapes in the $N$ and $R$ images, provided they are Nyquist sampled. However, errors in the PSF reconstruction can lead to flux misestimation and subtraction residuals.

Such issues arise when the PSF becomes strongly distorted, e.g. due to wind shake, out-of-collimation aberrations, or poor focus. In these cases, complex PSF morphologies may not be fully captured by the reconstruction, leading to residual artifacts, particularly around bright sources. One contributing factor is the finite size of the PSF stamp. Truncated PSF wings can bias the reconstruction and produce structured residuals as shown in Figure~\ref{fig:WindSubAndPSF}. In the current implementation, we adopt a $15\times15$ pixel PSF stamp as a compromise between capturing extended PSF structure and maintaining computational efficiency. Additionally, strong spatial variations of the PSF across the field may not be fully described by a single PSF model, further contributing to subtraction residuals.

To quantify the PSF reconstruction quality, we compute the following diagnostic measures:

\begin{description}
    \item [$\langle x_N^2\rangle,\langle y_N^2 \rangle, \langle x_N\times y_N \rangle$:] second moments of $P_N$
    \item [$\langle x_R^2\rangle,\langle y_R^2 \rangle, \langle x_R\times y_R \rangle$:] second moments of $P_R$
    \item [$\langle x_D^2\rangle_i,\langle y_D^2 \rangle_i, \langle x_D\times y_D \rangle_i$:] second moments of each source in the $D$ image
    \item [$\chi^2_{N,i}/\text{d.o.f.}$:] $\chi^2$ per degrees of freedom (d.o.f.) for the forced photometry fit in $N$ on the position of each source in the $D$ image
    \item [$\chi^2_{R,i}/\text{d.o.f.}$:] $\chi^2$ per degrees of freedom for the forced photometry fit in $R$ on the position of each source in the $D$ image
    \item [$\chi_{D,i}^2/\text{d.o.f.}$:] $\chi^2$ per degrees of freedom for the forced photometry fit in $D$ on the position of each source in the $D$ image
    \item [$\tilde{\chi}^2_{N,m}/\text{d.o.f.}$:] median $\chi^2_N/\text{d.o.f.}$ per magnitude bin in the $N$-image catalog
    \item [$\tilde{\chi}^2_{R,m}/\text{d.o.f.}$:] median $\chi^2_R/\text{d.o.f.}$ per magnitude bin in the $R$-image catalog
\end{description}

The second moments of $P_N$ and $P_R$, together with the PSF stamp size, help estimate the expected contamination from bright sources beyond the stamp boundary. The reduced $\chi_{N,i}^2$, $\chi_{R,i}^2$, $\chi_{D,i}^2$ values assess how well the PSF models fit each transient candidate-coincident source in $R$, $N$, and $D$ images, respectively. However, a caveat is that blended transients are not well represented by PSFs derived from isolated sources. Therefore, the $\tilde{\chi}_{N,m}^2/\mathrm{d.o.f.}$ and $\tilde{\chi}_{R,m}^2/\mathrm{d.o.f.}$ values are computed by binning all the sources in $N$ and $R$ by magnitude and taking the median per bin. A blended transient candidate of certain $N$- and $R$-image magnitudes can then be compared against the corresponding $\tilde{\chi}_m^2/\mathrm{d.o.f.}$ values to determine if the PSFs are well reconstructed within the particular magnitude ranges. These parameters are employed in several of the filtering steps, described below.

\subsection{Flagging for false alarms}
\label{sec:flagging}

The initial list of level 0 candidates largely consists of false-positive events that need to be separated from true positive events. Here we employ a set of classification filters based on hypothesis tests and heuristic methods. The filters do not remove candidates but assign bit-mask flags indicating likely false positives. No candidates are discarded at this step.
Table~\ref{tab:TransientsFilersBitMask} gives a summary of each individual filter and its associated bit-mask index. The filters are calibrated from real data on known asteroids and variable stars. A detailed description of each filter is given below.

\begin{table*}
\centering
\caption{The bit-mask dictionary of transients' filters}
\label{tab:TransientsFilersBitMask}
\begin{tabular}{lll}
\hline\hline
Bit name & Index & Description \\
\hline
Negative        & 0 &  The candidate has a negative S value \\
PSFChi2         & 1 &  $P_N$ or $P_R$ do not fit the candidate in the $N$ or $R$ image, respectively \\
Saturated       & 2 &  The candidate is near saturated pixels in the $N$ and $R$ images\\
BadPixelHard    & 3 &  The candidate is near bad pixels within the hard criteria \\
BadPixelSoft    & 4 &  The candidate is likely caused by an elevated single-pixel noise \\
StarMatch       & 6 &  The candidate matches a known star \\
MPMatch         & 7 &  The candidate matches a known minor planet (asteroid or comet) \\
Ringing         & 8 &  The candidate is caused by a ringing artifact \\
Overdensity     & 9 &  The local candidate density is above an empiric threshold \\
PSFShape       & 10 & The candidate is likely due to contamination in an $N$ image with a poorly reconstructed PSF shape\\
LIMMAG          & 11 & The candidate's $D$-image magnitude exceeds the limiting magnitude of the $N$ image \\
PVDist          & 12 & The candidate is caused by significant source displacement\\
Translient      & 13 & The candidate is caused by sub-pixel source movement rather than by source variability \\
Scorr           & 14 & The candidate is not significant when corrected for source noise \\
Streak          & 16 & The candidate is caused by a streak \\
Variable        & 18 & The candidate matches an already known variable object \\
NuclearNoise    & 19 & The candidate is caused by elevated noise of a galaxy nucleus\\
DiffSpike       & 21 & The candidate is caused by a diffraction spike\\
Extended        & 22 & The candidate is caused by an extended source or structure\\
\hline
\end{tabular}
\end{table*}

\begin{description}
    \item[\textbf{Bad pixels.}] Candidates are flagged for bad pixels based on two criteria: hard (BitIndex 3) and soft (BitIndex 4). The hard criterion suggests a clear fault near the transient candidate. Such faults are identified by querying the $N$, $R$, and $D$ bit-mask images within a $3$-pixel radius of the candidate position. A candidate is flagged if any of the nearby pixels match a bit-mask value containing a positive bit corresponding to:
    \vspace{4pt} 
    \begin{description}[itemsep=3pt]
        \item [\texttt{NaN}.] The pixel value is NaN
        \item [\texttt{Negative}.] The pixel value in $N$ or $R$ is negative
        \item [\texttt{Interpolated}.] The pixel value is interpolated
        \item [\texttt{Hole}.] The pixel is part of a ``hole" in the background of the $N$ or $R$ images
        \item [\texttt{NearEdge}.] The pixel is near an $N$, $R$, or $D$ image edge
    \end{description}
    \vspace{4pt}

\begin{figure}
    \centerline{\includegraphics[width=0.45\textwidth]{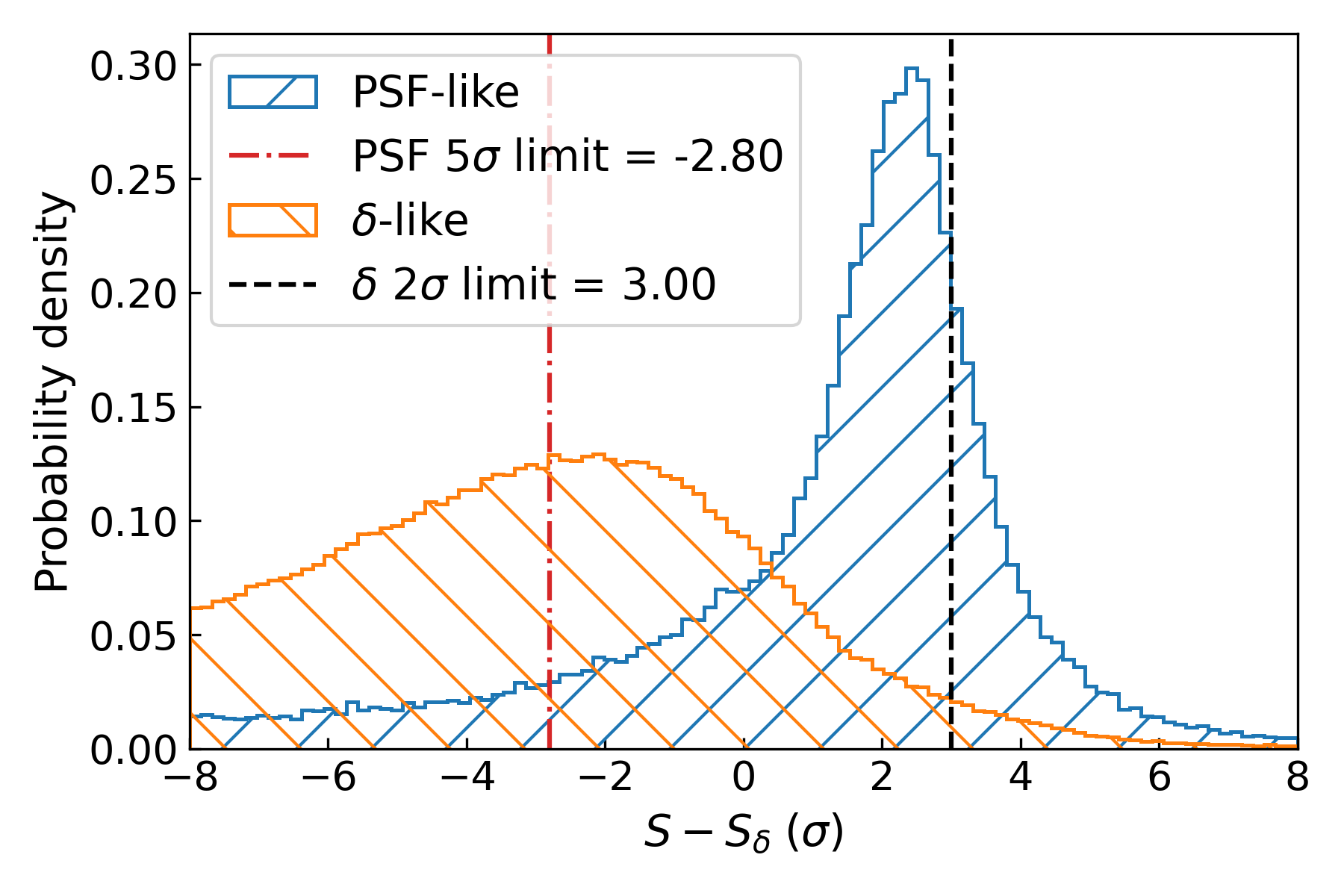}}
    \caption{Distributions of the difference between $S$ and $S_\delta$ for unblended residuals. The $\delta$-like distribution consists of $\sim5\times10^5$ residuals matching a positive \texttt{DarkHighVal} or \texttt{CR\_DeltaHT} bit, both indicative of hot single-pixel noise. The PSF-like distribution consists of $\sim10^5$ residuals not associated with either bit, and well fit PSFs $\chi^2/\mathrm{d.o.f.}<2$. Only $S>0$ residuals have been selected for both samples. Hot pixels produce false positive residuals characterized by a $\delta$-like signal $S-S_\delta < 0$. The red dash dotted line indicates the threshold above which residuals are retained by default. The dashed black line indicates the heightened threshold for residuals matched to the \texttt{DarkHighVal} or \texttt{CR\_DeltaHT} bits.}
\label{fig:BPSoftDists}
\end{figure}
    
    The soft criterion suggests that the significance of a candidate may be overestimated due to additional single-pixel noise components or cosmic-ray hits. As these are inherently single-pixel effects, they can be identified by comparing the statistic score $S$ derived using the $D$-image PSF with the score $S_\delta$ derived using a $\delta$-function kernel. Candidates for which $S-S_\delta < \epsilon_\delta$ are flagged as bad-pixel candidates. The default value of $\epsilon_\delta$ is set empirically from the archival distribution of unblended residuals that are not associated, within a $3$-pixel radius, with the following bits:
    
    \vspace{4pt} 
    \begin{description}[itemsep=3pt]
        \item [\texttt{DarkHighVal}.] Pixel bias or dark value is $>2\times$mean
        \item [\texttt{CR\_DeltaHT}.] Cosmic ray identified using hypothesis testing to $\delta$ function within a coadded image
    \end{description}
    \vspace{4pt}
    
    These bits indicate elevated single-pixel noise in the $N$ or $R$ image, which is propagated to a dominantly negative $S-S_\delta$ distribution (Figure~\ref{fig:BPSoftDists}). We compute the full width at half maximum (FWHM) of the PSF-like distribution and estimate the left-hand $5\sigma$ limit as the default $\epsilon_\delta$. For candidates associated with either bit, the threshold is increased to the right-hand $2\sigma$ limit of the $\delta$-like distribution to retain potential transients that may be spuriously coincident with such pixels. The $\sigma$-limits are chosen to be permissive towards potentially real transient candidates.
    
    A caveat of this filter is that a $\delta$-function feature in the $N$ or $R$ image does not necessarily propagate to a $\delta$-like feature in the $D$ image for high-count single-pixel events, due to artifacts introduced in the Fourier transformation. However, such extreme cases are effectively rejected by PSF $\chi^2$ tests instead.

    \item[\textbf{Saturation.}] The bit-mask images of $N$ and $R$ are queried for saturation within $3\,$pix of the candidate position. If there are pixels that are saturated in $N$ and $R$, the transient candidate is flagged for saturation (BitIndex 2). The candidate is not flagged if only one of the images shows saturation. 
    \item[\textbf{Ringing.}] The ZOGY statistic $S$ is compared against the Gabor-filter statistic $S_{\rm{g}}$. Candidates for which $S_{\rm{g}} > S$ are flagged as ringing artifacts (BitIndex 8).

    \item[\textbf{Diffraction spikes.}] Candidates near saturated sources are additionally tested for residuals aligned with the source center. Saturated regions are identified from the bit-mask image and grouped into connected components, from which centroids are computed. For each candidate near saturation, we draw a line between the candidate position and nearby saturation centroids and sample the $D$ image along that line. If a sufficiently large fraction of the sampled pixels show significant signal the candidate is flagged as a diffraction-spike artifact (BitIndex 21). This filter is intended to identify subtraction residuals that lie along spike-like structures extending from saturated stars.

\begin{figure}
\centerline{\includegraphics[width=0.45\textwidth]{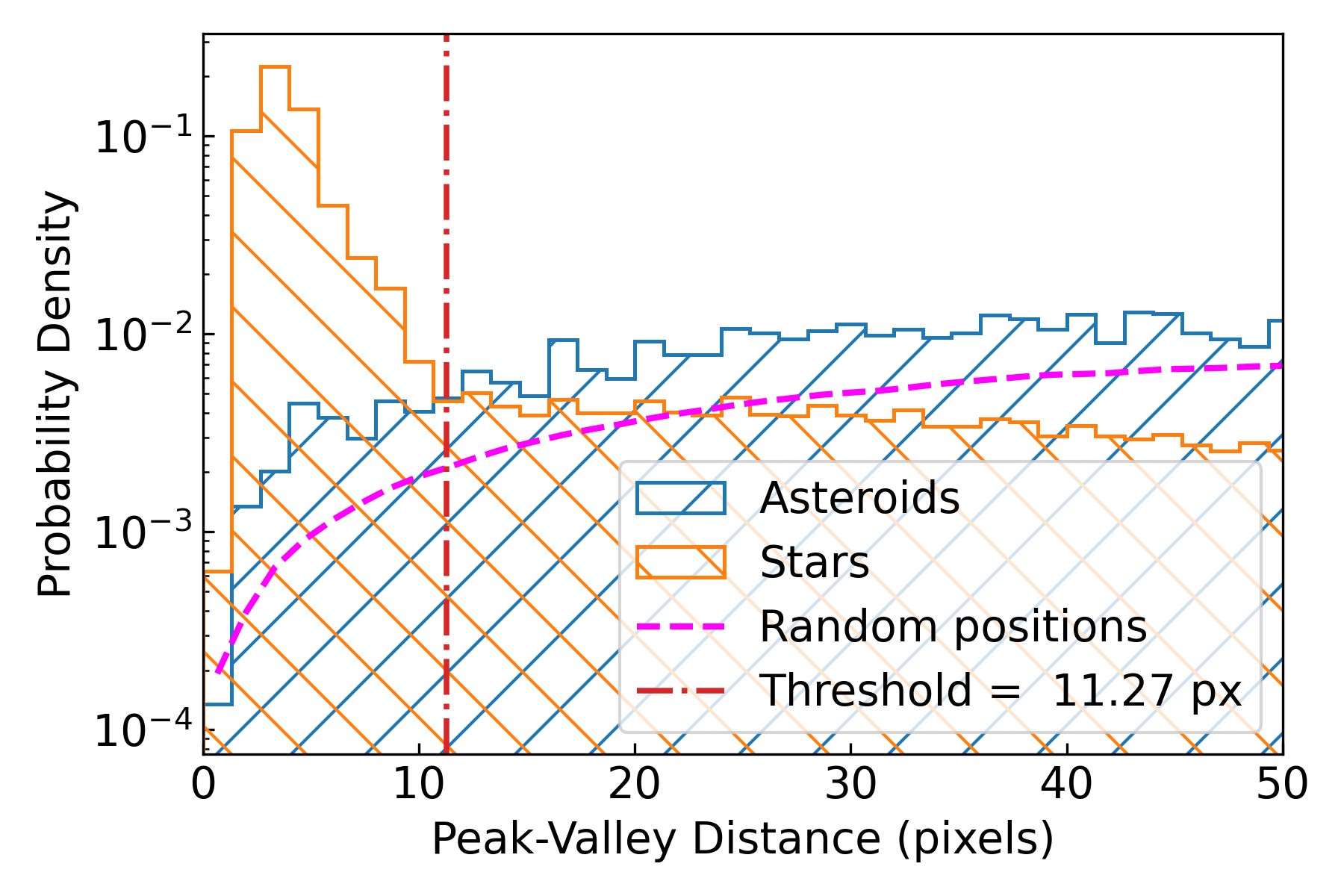}}
    \caption{Distribution of distances between nearest residuals of opposite sign in $S$. Shown are residuals matched to stars and asteroids, along with a spurious-association model (dashed magenta line). Positional shifts between the $N$ and $R$ images produce positive–negative residual pairs, leading to an excess at small separations in the star-associated sample. This excess is absent for asteroid-associated residuals, which follow the spurious-association model more closely. The red dash-dotted line indicates the threshold above which transient candidates are retained.}
\label{fig:PVDistDist}
\end{figure}
    
    \item[\textbf{Source displacement.}] A shift in the centroid of a persistent source between the $N$ and $R$ images generates a characteristic dipole residual, consisting of a positive peak and a corresponding negative valley in $S$. To identify such cases, we compute the distance $d_\text{PV}$ between each peak ($S \geq 5\sigma$) and its nearest valley ($S \leq -5\sigma$), and vice versa.
        
    For randomly distributed residuals, these associations are dominated by chance alignments, while for displaced sources, the peak–valley pairs are physically linked and occur at small separations. This produces a distinct excess at low $d_\text{PV}$ in the distribution of residuals associated with stars, which is not present for asteroid-associated residuals that follow a nearly random association model (Figure~\ref{fig:PVDistDist}). The random-association model is constructed by sampling random pixel positions within the image boundaries and measuring the distance to the nearest opposite-sign residual in the $D$-image catalog.
        
    This method is effective for non-subtle positional shifts, where both components of the dipole residual are detected. The threshold on $d_\text{PV}$ is determined empirically by comparing the distributions for star- and asteroid-associated residuals, and selecting a value that optimally rejects the former and retains the latter. Candidates with $d_\text{PV}$ below this threshold are flagged as false positives (BitIndex 12). This filter also captures artifacts that produce correlated positive and negative residuals, such as severe ringing and PSF reconstruction errors.

\begin{figure}
    \centerline{\includegraphics[width=0.45\textwidth]{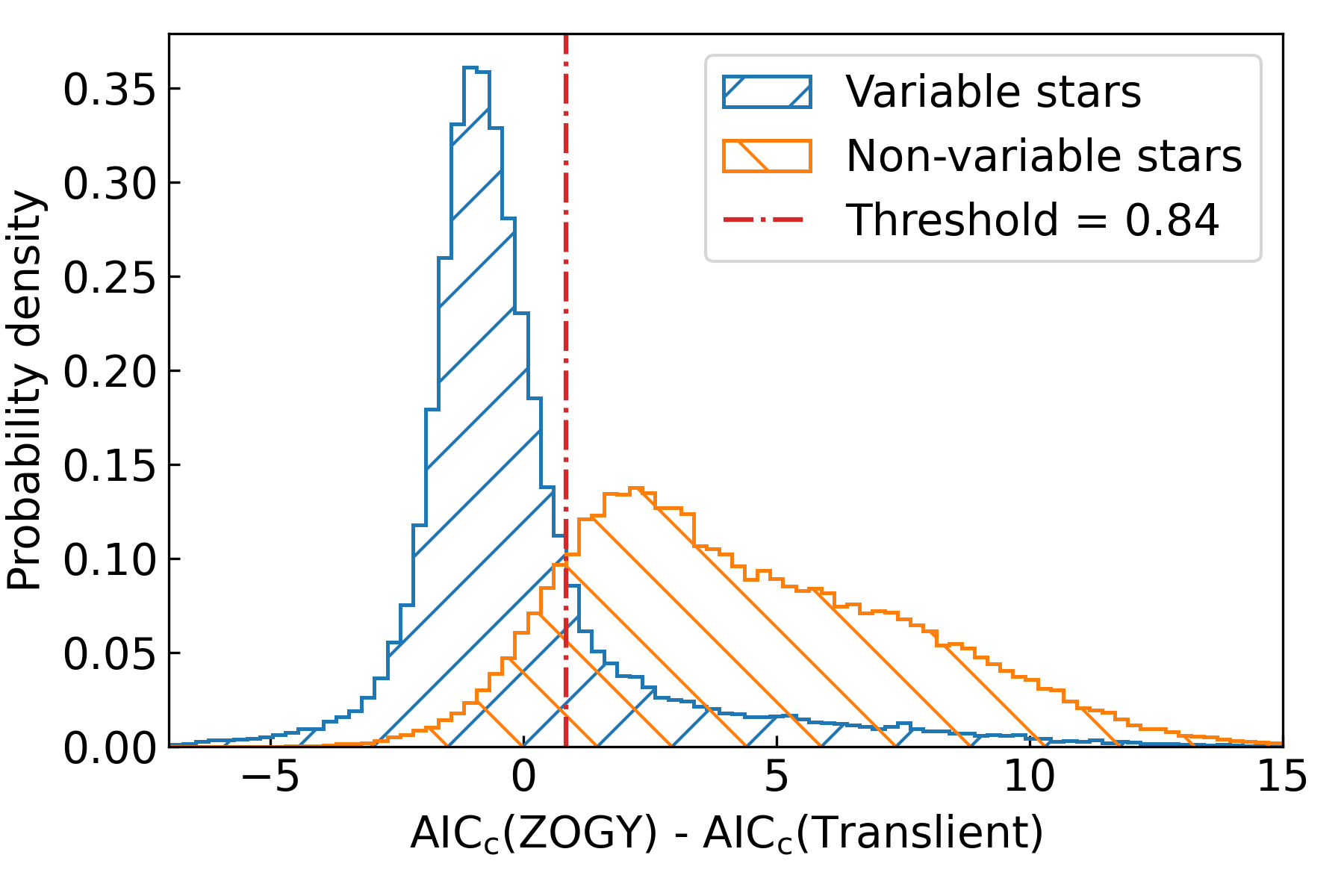}}
    \caption{Distribution of the difference between AIC$_\text{c}$(ZOGY) and AIC$_\text{c}$(\textsc{Translient}) for variable and non-variable stars. The red dash-dotted line indicates the selection threshold that maximizes the retention of variable-like candidates while minimizing contamination from non-variable-like candidates. Both shown distributions are restricted to residuals with $d_\mathrm{PV} > 11.3~$px, i.e., no opposite-sign residual is detected.}
\label{fig:AIC_diff}
\end{figure}
        
    \item[\textbf{Translient.}] For smaller positional shifts, where the opposite-sign counterpart may not be detected and the dipole becomes incomplete, the $d_\text{PV}$ approach loses sensitivity. In this regime, we compare the ZOGY statistic to the \textsc{Translient} statistic via the AIC$_\text{c}$ (\S\ref{sec:Translient}). If the \textsc{Translient} AIC$_\text{c}$ score is lower than the ZOGY AIC$_\text{c}$ score by a threshold value, the candidate is flagged as a false positive (BitIndex 13).
    
    The threshold is derived by comparing the AIC$_\text{c}$ differences between known variable and non-variable stars (Figure~\ref{fig:AIC_diff}), selecting the value that maximizes the retention of variable-like candidates while minimizing contamination. All candidates with $\vert S/N \vert < 3\sigma$ in forced photometry in the $R$ image automatically pass this filter.
    
    While the \textsc{Translient} statistic provides a general test for positional shifts and performs well for small shifts, its discriminating power decreases at larger separations. In this regime, the $d_\mathrm{PV}$-based approach provides a more robust and discriminating measure.
    
\begin{figure}
\centerline{\includegraphics[width=0.45\textwidth]{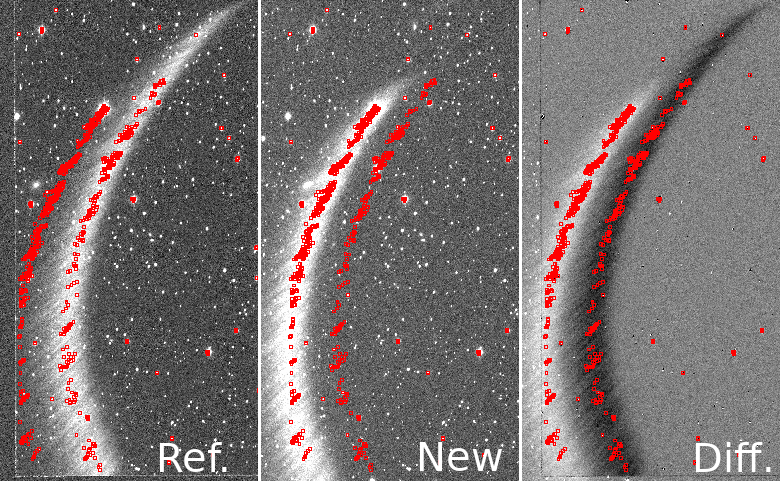}}
    \caption{Example of a large-scale flux elevation caused by a ghost. The ghost shifts between observations and creates a dense population of false positives (red markers) in the difference image.}
\label{fig:GhostDens}
\end{figure}

    \item[\textbf{Candidate density.}] Effects such as ghosts or stray light may elevate the intensity in a localized part of the camera over several $10$s to $1000$s of pixels. This region may then effectively move during the registration process, as well as due to changes in the telescope-to-sky or telescope-to-ground alignment during observations. These effects are generally difficult to characterize, but they tend to produce clusters of densely located false positives (Figure~\ref{fig:GhostDens}).
    
    To flag such candidates, the distances between all candidates are calculated, and the number of neighboring candidates within $100~$pix is counted. Since some neighboring candidates may already be explained by effects identified by other filters, only candidates that do not fail the bad pixel, ringing, \textsc{Translient}, and streak filters are counted as neighbors. This prevents the density metric from being dominated by false positives that are already classified by other criteria, and reduces the risk of rejecting real transients that happen to lie near such regions. Candidates matched to stars are also excluded as neighbors in order to avoid flagging transients in intrinsically crowded stellar regions.
    
    The neighbor density is then calculated as the sum of the inverse distances to all neighbors. A simple threshold on either the number of neighbors or the local density alone is insufficient, as real transients may occur in regions with moderate contamination or may have a single very nearby neighbor. In contrast, artifact-driven regions typically produce both a large number of neighbors and a high local concentration. We therefore adopt a combined metric that captures both aspects.
    
    All candidates are thus flagged when the product of the number of neighbors and the density exceeds a threshold value (BitIndex 9). The default threshold is $1.0$. Using asteroids as a proxy for transients, the chosen threshold provides a retention rate of $98.71\%$ (Figure~\ref{fig:NNeighDens}).

\begin{figure}
\centerline{\includegraphics[width=0.45\textwidth]{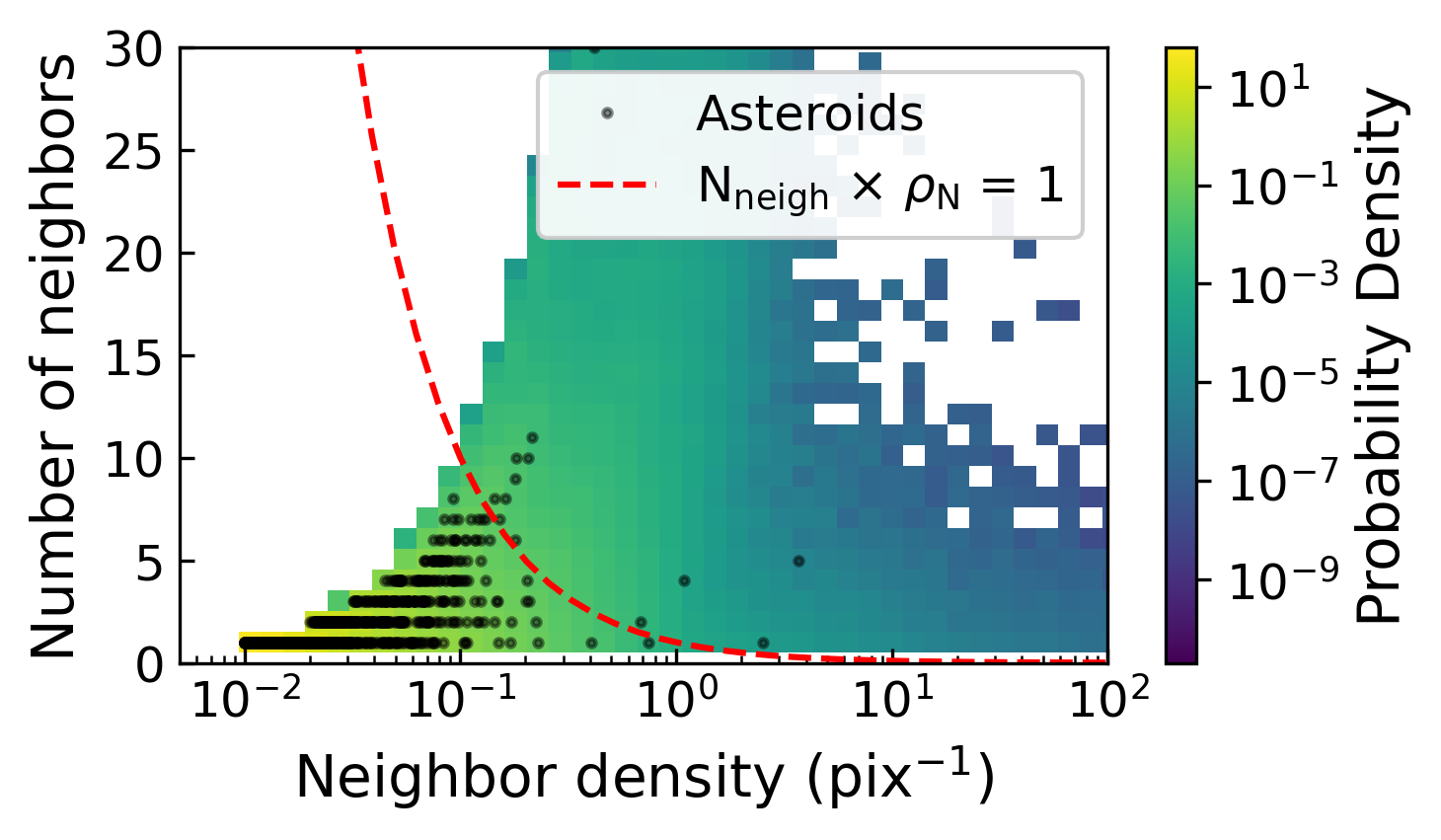}}
    \caption{Distribution of the number of neighbors $N_{\text{neigh}}$ and the neighbor density $\rho_{\text{N}}$. Neighbors are counted within a $100$ pix radius, and $\rho_{\text{N}}$ is defined as the sum of the inverse distances to all neighbors, $\rho_{\text{N}} = \sum_{i=1}^{N_{\text{neigh}}} d^{-1}_i$, where $d_i$ is the distance to the $i$th neighbor. Shown is the distribution for $\sim40$ million subtraction residuals. Residuals coincident with asteroids are shown as black dots. A threshold of $N_{\text{neigh}} \times \rho_{\text{N}} = 1$ is indicated by the red dashed line.}
\label{fig:NNeighDens}
\end{figure}
   
    \item[\textbf{PSF shape.}] As discussed in \S \ref{sec:PSF_reco_errors}, the subtraction process is sensitive to the quality of PSF reconstruction. The filtering therefore combines global assessment of the PSF quality with local tests at each candidate position (BitIndex 10). We first evaluate the PSF quality using the second moments of $P_N$. Images with small second moments are considered well-behaved and are treated more permissively, while larger second moments indicate broader or distorted PSFs, for which stricter thresholds are applied. This selective approach allows for some tolerance in otherwise problematic images, while still excluding clear subtraction artifacts.
    
    For each candidate, we estimate the level of contamination from nearby persistent sources as follows:
    \begin{enumerate}
        \item Retrieve positions and fluxes $I_R$ of all persistent sources from the $R$-image catalog.
        \item Estimate the $N$-image fluxes as
        \begin{equation}
            I_N=I_R\times10^{0.4 (Z_N-Z_R)},
        \end{equation}
        where $Z_N$ and $Z_R$ are the magnitude zero points of the $N$ and $R$ images, respectively.
        \item Estimate the tail flux beyond the PSF stamp of all the persistent sources in $N$ as 
        \begin{equation}
        I_{N,T} = I_N\left(1-\mathrm{erf}\left(\frac{L_{P_N,x}/2}{\sqrt{2\langle x_N^2\rangle}}\right)\times         \mathrm{erf}\!\left(\frac{L_{P_N,y}/2}{\sqrt{2\langle y_N^2\rangle}}\right)\right),
        \end{equation}
        where $L_{P_N,x}$ and $L_{P_N,y}$ are the sizes of the PSF stamp in the $x$ and $y$ direction, respectively.
        \item Count all the persistent sources with $I_{N,T}$ exceeding a fraction of the background rate as contaminators.
        \item Match candidates to nearby contaminating sources within a radius that scales with the PSF size, excluding very small separations to avoid self-contamination.
        \item For matched candidates, compare the candidate flux $I_D$ to the summed contamination flux $\Sigma _i I_{N,T,i}$. Candidates that are sufficiently bright relative to this contamination are retained.
    \end{enumerate}
    In addition to this contamination test, we require that the local subtraction environment is reasonably clean, based on annulus statistics in the $D$ image.
    The contaminator positions are taken from the $R$-image catalog in order to avoid bias from transients present in the $N$ image and to mitigate centroid shifts caused by blended sources, e.g., a supernova near a galaxy nucleus. This ensures that the contamination estimate reflects only persistent sources.

    \item[\textbf{PSF-fit photometry.}] For each candidate, the quality of PSF reconstruction is additionally tested via the $\chi^2/\text{d.o.f.}$ of the PSF fit on $N$ and $R$ images (BitIndex 1). 

    If the candidate is blended with a source that is present in $N$ and $R$ images, the test is performed against the median values $\tilde{\chi}^2_{N,m}/\text{d.o.f.}$ and $\tilde{\chi}^2_{R,m}/\text{d.o.f.}$ derived from the respective native source catalogs. If the candidate's respective median value is outside a threshold interval, it is flagged as a false positive due to poor PSF reconstruction in the given magnitude range. The default interval is $[0.1,1.2]$.
    
    For non-blended candidates, the $\chi^2/\text{d.o.f.}$ test is applied to the candidate's direct fit result, rather than to a global median. In this case, the $R$ image position contains only background, and only a test on $\chi_{N,i}^2/\text{d.o.f.}$ is required. The default interval is $[0.1,2.0]$. This step furthermore filters artifacts such as a few pixel-long streaks caused by bright stars in flat-field images, which are taken during twilight.

    \item[\textbf{Extended sources.}] $S$ is compared against the extended-source significance estimator $S_{\rm Ext}$ to identify candidates that are broader than the PSF. Such residuals are inconsistent with point-like transients and may arise either from subtraction artifacts or from genuinely extended astrophysical sources such as comets. Since neither case corresponds to a point-like transient, these candidates are filtered. Specifically, candidates are flagged as extended (BitIndex 22) when $S - S_{\rm ext} < \epsilon_{\rm ext}$, where $\epsilon_{\rm ext}$ is estimated as the lower 5th percentile of residuals on variable stars ($\sim-0.25$).
    
    \item[\textbf{Negative transients.}] New transients mostly appear as positive residuals in the $D$ image and thus produce a positive $S$ value. However, negative $S$ values may also be physical, corresponding to fading sources. For new-transient discovery, we flag negative-$S$ candidates (BitIndex 0).
    
    \item[\textbf{Source noise.}] Candidates with a source-noise corrected statistic $\vert S_\text{corr}\vert<5\sigma$ may be flagged as false positives (BitIndex 14). However, the source noise of faint transients may be overestimated, which will over-correct the transient's significance and filter it as a false positive. This may occur due to, e.g., a heightened local background rate. Thus, candidates for which $S-S_\text{corr}<\delta_\text{corr}$ also pass this filter. The default $\delta_\text{corr}$ is $0.7$.
    
    \item[] \textbf{Matching with stars.} All candidates are matched against the GAIA~DR3 catalog \citep{GAIA+2022yCat_GAIA_DR3_MainSourcesCatalog} to identify coincidence with known stars. Bright stars are a common source of artifacts, e.g., due to saturation, increased noise, or deviations from i.i.d.\ noise assumptions. In addition, airmass-dependent color terms between the $N$ and $R$ images (\S \ref{sec:flux_ratio_unc}) can produce spurious variability. We therefore assign each candidate a stellar association score based on positional consistency and GAIA astrometric significance, and flag candidates with sufficiently high stellar probability as false positives (BitIndex 6).
    
    To avoid misclassifying galaxy-associated residuals as stellar, all candidates are also matched against the GLADE+ \citep{Dalya+2021_GLADEplus_GalaxyCatalog} and PGC \citep{Paturel+1989_HyperLeda_PGC_Galaxy_Catalog} catalogs. When both stellar and galaxy associations are present, the classification is based on their relative probabilities, and candidates consistent with galaxy association are not flagged as stellar. In the future, with improved characterization of problematic stars and better treatment of color terms, this filter may be relaxed to enable the detection of genuine stellar transients.
        
    \item[\textbf{Nuclear noise.}] Regarding transient detection, galactic nuclei present a similar challenge as stars. To still detect and report nuclear transients, nuclear candidates are subjected to further testing. All the candidates within a sub-image are matched to the $R$-image catalog, and a list of coincident persistent sources is retrieved. This gives an association between the $R$-image magnitude and the $S$ score for residuals on persistent sources, including galactic nuclei. For each nuclear candidate, comparison candidates are selected that are brighter by up to $0.5~$mag in the $R$-image. This ensures that the nuclear candidate is the faintest source in the comparison bin, and, if it is not variable, it is expected to have the lowest $S$ score. If the candidate is a transient, it is expected to have an $S$ score above the median of the selected candidates. Thus, the nuclear candidate is flagged if its $S$ score is below the $68$th percentile $S$ score of its respective $R$-image magnitude bin (BitIndex 19). If no other candidates coincide with known $R$-image sources within the magnitude bin, then the subtraction is assumed to be free of residuals, and the nuclear candidate is retained. Bright nuclei ($<17\,$mag) must exceed a stricter threshold, with their $S$ score falling above the $95$th percentile of the comparison bin.
    \item[\textbf{Matching with Solar System objects.}] Candidates are matched against known numbered and unnumbered minor planets with a predicted magnitude brighter than $21.5~$mag. The matching is done by numerically integrating the orbits of all 
    known minor planets\footnote{Implemented in: {\tt imProc.match.match2solarSystem}.} and matching with an angular radius of $10''$. The distance to the nearest minor planet and its magnitude in the $N$ and $R$ images are recorded. All the 
    candidates matched to a minor planet in the $N$ or $R$ image are flagged (BitIndex 7).
    \item[\textbf{Matching with variable sources.}] Candidates that have been matched to GAIA~DR3 sources within a $3''$ distance are additionally probed on whether the matching sources are variable (BitIndex 18). Here, the \textit{phot\_variable\_flag} argument within the GAIA~DR3 catalog is used. All candidates are additionally matched to the Million Quasars catalog \citep{Flesch2015_TheMilionQSO_Catalog} to identify candidates caused by active galactic nuclei (AGNs). Candidates coincident with variable GAIA~DR3 sources or AGNs are flagged as known variable sources.

\end{description}

The flagging process appends an additional column to the level $0$ catalog with a bit-mask value summarizing all non-passed filters. The bit-mask value of the candidates that pass all filters is $0$. The list of the candidates after single-epoch flagging is referred to as the level $1$ catalog and is saved to disk. Level 1 candidates are then sub-selected for database injection and for multi-epoch matching. The sub-selection is performed in order to maintain the growth of the transients database at a reasonable pace. The candidates that fail a few specific filters are thus removed from the catalog before injection. The filters specified for this purpose are currently \texttt{BadPixelHard}, \texttt{LIMMAG}, \texttt{Negative}, \texttt{Overdensity}, \texttt{PVDist}, \texttt{Streak}, and \texttt{PSFShape}, see also Figure~\ref{fig:SequantialAndIndividualFilters} (top). In the near future, this cleaning process will be refined with more conditional selection in order to, e.g., retain strongly dimming sources that fail the \texttt{Negative} filter. The sub-selected list is referred to as a level $2$ catalog. The collection of saved parameters per candidate at level $2$ is shown in Appendix~\ref{app:fullcatalog}. Following the injection of level $2$ candidates, a transient classification and merging process is initiated over the candidate database. The purpose of this process is to identify multi-epoch transients and to link single detections of unknown asteroids into arclets. Finally, new detections are reported to the IAU official Transient Name Server (TNS) \citep{TNS} and to the Minor Planet Center (MPC) \citep{MPC}. An online tool (Marshal) called Candidate Alert System for Transients (CAST) is being developed for inspecting and activating follow-up observations on these candidates. These processes will be described elsewhere.

\section{Performance}
\label{sec:performance}

A single $6$-h night of observations with a single LAST telescope produces on average $\sim 1.7$ million initial $\vert S\vert \geq 5\sigma$ transient candidates, as seen in early estimates. An extrapolation to $72$ LAST telescopes suggests that $>100$ million candidates will be produced by the full array per night.The vast majority are false positives, requiring high-purity filtering with minimal loss of real transients. Here we provide an evaluation of some of the early core performance metrics of the transient detection pipeline, including the limiting magnitude, the expected transient rates, the preliminary efficiency and purity estimates, and the runtime. These results demonstrate that the current implementation already provides a robust and operational transient detection system suitable for large-scale surveys, even prior to further optimization.

\subsection{Limiting magnitude}

The $5\sigma$-limiting magnitude of the transients detection pipeline for a PSF-like source can be estimated for each of the $D$ images as

\begin{eqnarray}
    M_\text{D,lim} &=& Z_D - 2.5\log\left(5\frac{\sigma_D}{\Vert P_D\Vert}\right),
\end{eqnarray}

\noindent where $Z_D$ is the $D$-image zero point in magnitudes, $\Vert P_D \Vert$ is the norm of the $D$-image PSF, and $\sigma_D$ is the $D$-image standard deviation. While $\sigma_D$ is the standard deviation per pixel, $\sigma_D/\Vert P_D\Vert$ gives the standard deviation of the match-filter process \citep{Zackay+2016_ZOGY_ImageSubtraction}. $M_\text{D,lim}$ depends on the depth of the $N$ and $R$ images and thus varies with the exposure time of both images. The exposure time of $N$ is $20\times 20\,$s for the nominal LAST survey strategy. The exposure time of the pre-generated $R$ images will, however, vary with the availability of good images for reference production for a particular part of the sky. 

\begin{figure}
\centerline{\includegraphics[width=0.45\textwidth]{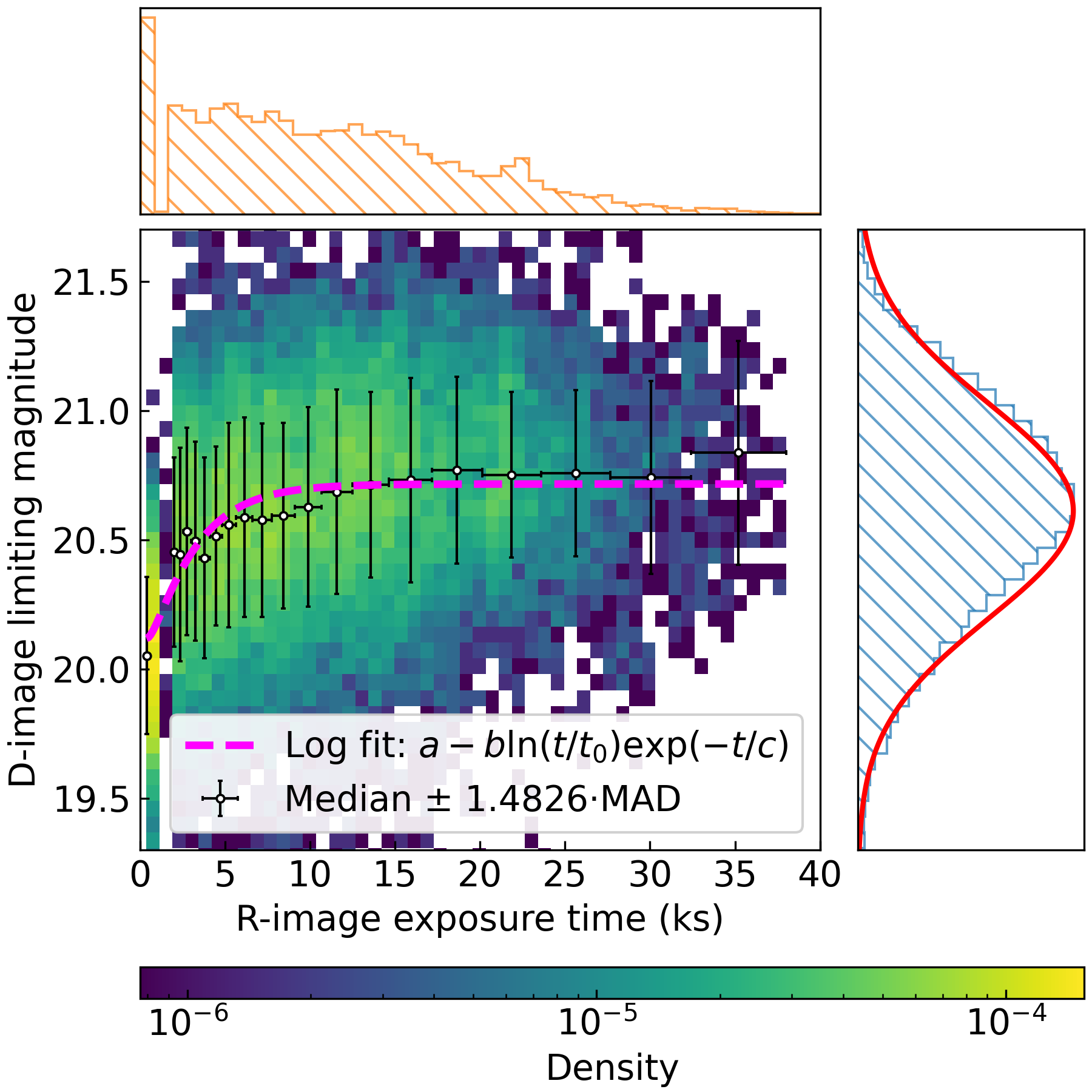}}
    \caption{Distribution of the $5\sigma$-limiting magnitude of the transient-detection process versus total $R$-image exposure time, using the most recent stacked $R$ images. The central panel shows a 2D density histogram. The top and right panels show the marginal distributions of exposure time and limiting magnitude. Black and white points mark per-bin medians from logarithmic exposure bins, and error bars are $1.4826\times\mathrm{MAD}$ per bin, where $\mathrm{MAD}$ is the median absolute deviation. The magenta dashed curve in the central panel is a fit to the medians, and the red solid curve in the right panel is a Gaussian fit to the overall limiting-magnitude distribution.}
\label{fig:DLimMagDeep}
\end{figure}

Figure~\ref{fig:DLimMagDeep} shows the distribution of the D-image limiting magnitudes as a function of the $R$-image exposure time $t_\text{R}$ using the most recent LAST $R$ images at the time of this writing. A Gaussian fit over the entire limiting magnitude distribution in this case provides an average of $(20.6 \pm 0.4)\,$mag. It is important to note that deeply stacked $R$ images cover several observation cycles and combine images over large variations in weather, lunar cycle, and instrumental issues. This naturally increases the scatter in the estimated limiting magnitude.

We derive the shallow end by fitting a Gaussian to the slice of $20\times20\,$s $R$-image exposure, which yields $(20.3 \pm 0.3)$ mag. With higher $t_{\rm R}$, we expect the $D$-image limiting magnitude to improve roughly logarithmically until it saturates at a plateau set by the fixed $N$-image exposure. We model this with
\begin{eqnarray}
  M_{\rm D,lim}(t_{\rm R}) &=& M_{\rm D,lim,deep} - A\,\ln\left(\frac{t_{\rm R}}{t_0} \right)\,\exp\left(-\frac{t_{\rm R}}{T_{\rm R,deep}}\right),
\end{eqnarray}
where $M_{\rm D,lim,deep}$ is the asymptotic plateau depth, $A$ is the amplitude of the logarithmic term, $T_{\rm R,deep}$ sets how quickly the plateau is approached, and $t_0$ is a reference time, here $t_0 = 1\,$s. 

To mitigate density bias and heteroscedastic variance across exposure times, we fit the model to binned medians. Specifically, the data are binned in logarithmic exposure-time bins. In each bin we compute the median limiting magnitude and its dispersion as $\sigma_i = 1.4826\times\mathrm{MAD}$, where MAD is the median absolute deviation. This dispersion equals the standard deviation for Gaussian data. We then perform a weighted fit to the per-bin medians with weights $w_i=1/\sigma_i^2$, and discard bins with fewer than $100$ entries. This yields $M_\mathrm{D,lim,deep}=(20.7\pm0.1)$ mag, $A=(0.12\pm0.07)$ mag, and $T_\mathrm{R,deep}=(2400\pm1800)$ s. The value of $M_\mathrm{D,lim,deep}$ is consistent with the depth expected when the $R$ image is effectively noiseless, that is, when the limiting magnitude is set by the fixed $N$-image exposure. The large uncertainty on $T_\mathrm{R,deep}$ likely reflects undersampling at $R$ exposures of a few kilo-seconds. Even so, the results suggest that exposures $\gtrsim5$ ks are needed before the gain from deeper $R$ stacks is largely exhausted.

\subsection{Transients rates}
\label{sec:TransientsRates}
We have applied the LAST Pipeline~II to a random subset of LAST observations to evaluate the flagging performance, estimate the expected transient rates, and compare these to statistical false alarm rates. 

\subsubsection{Dataset and setup}
The selected sample includes $\sim40$ million initial candidates, which is roughly $1/3$ the estimated nightly rate for 72 LAST telescopes. The selection was performed randomly on the data taken by LAST in July--August 2024. This time marks a commissioning phase when 24 telescopes participated in the all-sky survey at sub-optimal efficiency. Thus, the data span several observation nights. To exclude inhomogeneities within the sample potentially caused by different $R$-image exposures, only $20\times20\,$s $R$ images have been employed within this study.

\subsubsection{Expected statistical fluctuations}

Given the size of the sample, it is useful to estimate the expected amount of $S\geq5\sigma$ Gaussian fluctuations. Each single-telescope LAST observation produces 24 science $N$ images, derived from a single full-frame raw image. Each $N$ image has a pixel size of $1726\times1726$. The trial sample comprises $1304$ single-telescope observations and thus $1304\times24 = 31296$ subtractions. The correlation length across the image is given by the PSF, as it serves as the match-filter kernel. The Gaussian standard deviation of the median LAST PSF is $\sim1.46''$, or $\sim1.17\,$pix. This gives an order-of-magnitude estimate of $1726^2/(1.17^2\pi)\approx7\times10^5$ independent trials per subtraction. Assuming a Gaussian noise distribution, the probability of a single trial exhibiting a $\geq 5\sigma$ fluctuation is $2.87\times10^{-7}$, implying $\sim0.20$ expected spurious detections per subtraction and $\sim6\times10^3$ over the full dataset presented here. 

This estimate should be interpreted as an order-of-magnitude upper limit on the number of statistical fluctuations prior to filtering. These fluctuations are defined at the scale of the PSF and are therefore intrinsically PSF-like. Such fluctuations are thus statistically indistinguishable from real point sources in a single epoch, and cannot be fully suppressed by filtering without also rejecting real low-flux transients. This demonstrates that, at the scale of the survey, a $5\sigma$ detection threshold is intrinsically insufficient for reliable single-epoch transient identification, as thousands of spurious detections are excepted even in the absence of systematic effects.

\begin{figure}
\centering
\includegraphics[width=0.5\textwidth]{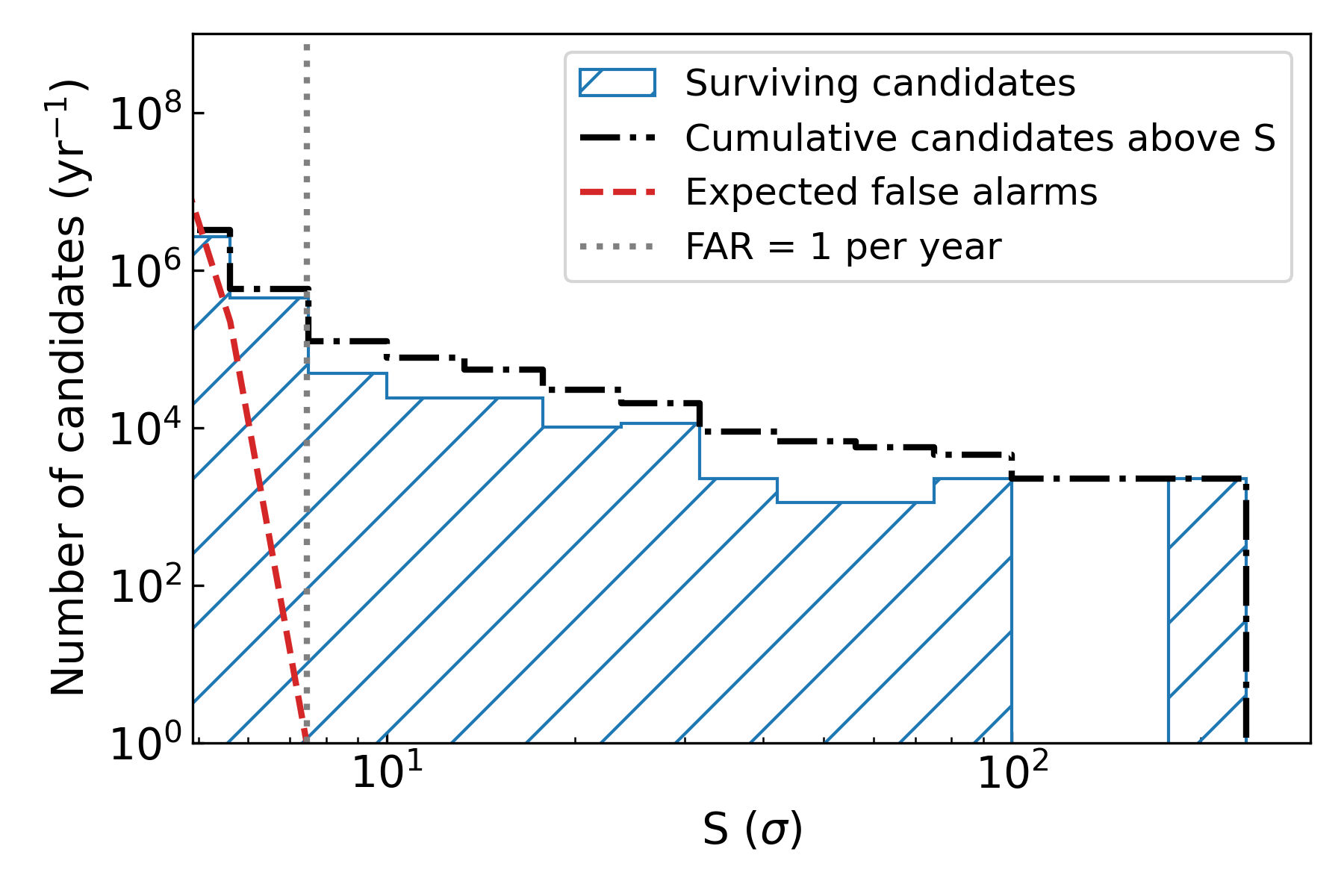}
\caption{Estimated number of surviving transient candidates per year for 72 telescopes and 6~h exposure per night. The cumulative rate above a given $S$ value is shown by the black dash-dotted line. The expected false alarm rate (FAR) from Gaussian statistics is shown by the red dashed line, and FAR$ = 1\,$yr$^{-1}$ is marked by a gray vertical dotted line.}
\label{fig:PassingCand_vs_FAR}
\end{figure}

\subsubsection{Empirical survivors and yearly rate}
\label{sec:yearly_rates}

Applying the single-epoch filtering pipeline to the $\sim40$ million initial candidates results in $2849$ surviving candidates with $S\geq5\sigma$, about half of the estimated number of unfiltered statistical fluctuations. Given the approximate nature of the fluctuation estimate, this level of agreement indicates that a substantial fraction of the surviving candidates can be attributed to statistical fluctuations alone.

In addition, the surviving population is expected to include systematic residuals that are not captured by the Gaussian noise model, such as those caused by imperfect subtractions and instrumental artifacts. As a result, single-epoch detections at the $\sim5\sigma$ level remain strongly affected by non-astrophysical residuals. This demonstrates that additional selection criteria are required to isolate real transients. In practice, this is achieved by imposing higher significance thresholds or by requiring multi-epoch detections, both of which strongly suppress the contribution from statistical fluctuations.

We can estimate the yearly rate of passing transient candidates as a function of $S$ by extrapolating the surviving candidate sample and comparing it to the expected false alarm rate (FAR) from Gaussian fluctuations. Figure \ref{fig:PassingCand_vs_FAR} shows both rates for $1\,$yr of observations with $72$ telescopes, assuming $6\,$h of observations per night. Considering only Gaussian fluctuations, the FAR drops below $1\,\mathrm{yr}^{-1}$ at $S\gtrsim7.5\,\sigma$. This FAR should be interpreted as an idealized, unfiltered upper bound set by statistical fluctuations alone, while the empirical candidate distribution reflects statistical, systematic, and astrophysical contributions. Extrapolating the surviving-candidate distribution predicts $>10^{5}$ single-epoch detections per year above $7.5\sigma$ to pass the filtering stage.

\subsubsection{Per-filter behavior}
The sequential effect of all single-epoch filters and the individual effect of each filter are summarized in Appendix~\ref{app:singlefilters}. In brief, the dominant sources of false positives are bright stars, poor PSFs or PSF reconstruction, single-pixel peaks such as cosmic ray hits and bad pixels, and image registration errors.

\subsection{Purity}

The 2849 surviving candidates are dominated by low-significance spurious detections, so further selection criteria are needed to identify real transients. Requiring multi-epoch detections, or $S\geq7.5\,\sigma$, in a single epoch for FAR$\,< 1\,$yr$^{-1}$ (see $\S$\ref{sec:yearly_rates}) reduces the set to $128$, corresponding to a surface density of $\sim0.01\,$deg$^{-2}$. Manual inspection of the $128$ surviving candidates (Table~\ref{tab:sample_classification}) shows that $\sim80\%$ are known or new real transients, $\sim8\%$ are Galactic variable or transient sources (CVs and stellar flares), $\sim5\%$ are extragalactic variable sources (AGNs, RR Lyrae variables), and $\sim8\%$ are false positive residuals. As Galactic transients may be counted as either signal or noise, depending on the scientific motivation, the interpretation of the purity within the sample may differ. With a focus only on new transient sources at time of detection and artificial bogus events, a purity of $\gtrsim90.0\%$ is estimated for this sample when considering only the $S\ge7.5\sigma$ candidates.

\begin{table}
\centering
\caption{Classification of surviving candidates from $\sim 40$ million initial candidates. 
Case A: single-epoch $S \ge 7.5\sigma$. 
Case B: single-epoch $S < 7.5\sigma$ with a coincident detection at another epoch.}
\label{tab:sample_classification}

\begin{tabular}{lcc}
\hline\hline
Classification & Case A & Case B \\
\hline
\\[-3pt]
Transient, TNS counterpart      & 76  & 18 \\
Transient, ZTF counterpart      & 3  & 0 \\
Transient, LAST only            & 5  & 0 \\
[1pt] \hline \\[1pt]
CV                              & 9  & 0 \\
M-dwarf flare                   & 1  & 0 \\
AGN                             & 3  & 2 \\
RR Lyrae variable               & 1  & 0 \\
[1pt] \hline \\[1pt]
Diffraction spike               & 0  & 2 \\
PSF wing residuals              & 6  & 0 \\
Under subtracted nuclei         & 2  & 0 \\
\hline
\end{tabular}

\end{table}

\subsection{Efficiency}

We estimate the single-epoch filtering efficiency as a function of $S$, i.e., the detection $S/N$, using simulated transients. We select random real science images and inject artificial sources near detected galaxies. Galaxies are identified by cross-matching the $N$-image source catalog against the PGC catalog \citep{Paturel+1989_HyperLeda_PGC_Galaxy_Catalog}. For each identified galaxy, we inject a source uniformly within its projected elliptical footprint defined by the PGC parameters. The semi-major axis is given by

\begin{eqnarray}
    a &=& 3 \times 10^{\mathrm{logD25}} \ \mathrm{arcsec} \\
    \mathrm{logD25} &=& \log_{10}(D_{25}/0.1\,\mathrm{arcmin}),
\end{eqnarray}

\noindent where $D_{25}$ is the apparent major-axis diameter measured at the $B = 25\,\mathrm{mag\,arcsec^{-2}}$ isophote. The semi-minor axis is

\begin{equation}
    b = a \times 10^{-\mathrm{LogAxisRatio}}.
\end{equation}

\noindent The ellipse is oriented according to the PGC position angle $\mathrm{PA1950}$. Injection positions are drawn uniformly in area within the ellipse by sampling

\begin{equation}
    \rho = \sqrt{u}, \qquad u \in [0,1], \qquad \phi \in [0, 2\pi),
\end{equation}

\noindent with $u$ and $\phi$ drawn from uniform distributions. The positions are first defined in the galaxy-aligned frame as

\begin{equation}
    x' = a \rho \cos\phi, \qquad y' = b \rho \sin\phi,
\end{equation}

\noindent and then rotated by $\mathrm{PA1950}$ to obtain sky offsets in east and north directions $(\Delta E, \Delta N)$, which are converted to equatorial coordinates $(\alpha, \delta)$ using a local tangent-plane approximation.

Each artificial source is injected with a flux corresponding to a PSF S/N of $[5,30]$ in the $N$ image. Sources are not injected into overlap areas, consistent with the removal of duplicate detections. We then execute the Pipeline II as normal and extract detected artificial transients from the $D$-image catalog.

\begin{figure}[tbp] 
\centering
\includegraphics[width=\columnwidth]{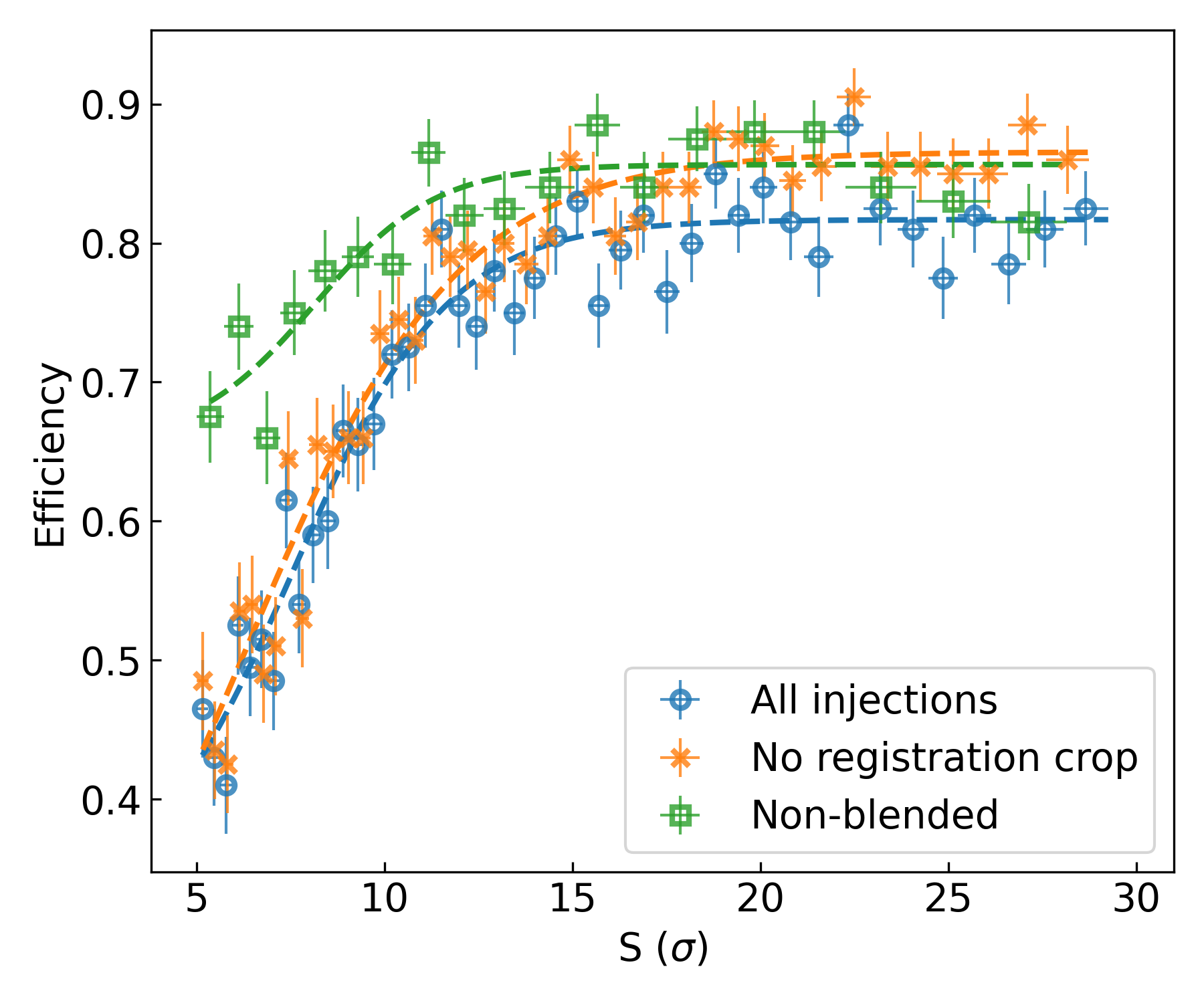}
    \caption{Estimated transients pipeline efficiency in relation to candidate significance. The estimate is derived for a sample of $10^4$ artificial transients (blue circles). Two sub-selections are shown, one for the injections outside of the cropped registration areas (orange x's), and one for injections not blended with any cataloged sources (green squares). Each bin consists of $200$ injections. Error bars indicate bin widths. A logistic sigmoid fit is shown for each sample.}
\label{fig:InjectionsEfficiency}
\end{figure}

Here we present a sample of $10^4$ injections. The rejection fraction per filter is shown in Fig.~\ref{fig:InjectionsIndividualFilters} in Appendix~\ref{app:singlefilters}, and the total survival fraction as a function of $S$ is shown in Figure~\ref{fig:InjectionsEfficiency}. Within the entire sample, $\sim\mathrm{73}\%$ of injections survive the filtering pipeline. The dominant losses occur in a few specific filters.

The \texttt{BadPixelHard} filter flags $\sim4.7\%$ of injections and requires a discussion. The dominant bit-mask values triggering the filter are \texttt{NearEdge} at $\sim1.61$\% and \texttt{NaN} at $\sim3.34$\%. Assuming perfect overlay between images, one would estimate these rates to be near zero. In particular, given that we inject sources outside of the $64$ pix overlap border, no sources are near the image edge and as such should not trigger the filter via the \texttt{NearEdge} bit. However, pointing offsets between observations reduce the viable FoV for transient detection. All of the candidates flagged due to the \texttt{NaN} bit were matched to a \texttt{NaN} pixel in the $R$ image. In the case of the \texttt{NearEdge} bit, $\sim73$\% were matched to a \texttt{NearEdge} marked pixel in the $R$ image, and $\sim26$\% in the $N$ image. The contribution from the $N$ image is non-zero because the single $20~$s images within a $20\times20~$s coadd may have significant shifts, whereas the bit masks are propagated via the logical OR operator during coaddition.

However, $\sim4.5$\% of injections are flagged due to misalignment between the $N$ and $R$ images in which the FoV is effectively truncated (see \S\ref{sec:Registration}). Furthermore, reference-misalignment within a given sub-image partition also likely extends to misalignment in the neighboring sub-images, reducing the retention of duplicated detections in overlap region, which has been ignored in these simulations. In a similar estimation of the efficiency using asteroid detections, we find a total rejection rate of transients up to $\sim9$\% due to the $N$-$R$-misalignment. Albeit this rate is inflated by spurious associations between asteroids and the high rate of artifacts in such regions. A conservative estimate for the true rejection is thus 5--9\%. Such a high rejection rate clearly shows the inefficiency of the current reference scheme. Works are thus underway to enhance the LAST pipeline with a field-independent reference scheme (Krassilchtchikov et al., in prep) that will reduce these masking effects in future datasets, including under arbitrary positions and orientations of the $N$ image. To estimate the efficiency with improved overlay between images, we consider only the subset of injections that pass the \texttt{BadPixelHard} filter. In this case, the overall survival rate increases to $\sim76\%$. An efficiency curve for a sub-selection of injections with no registration cropping is shown in Figure~\ref{fig:InjectionsEfficiency}.

Among the remaining filters, the ones with the rejection rate exceeding $\gtrsim5\%$ are \texttt{PSFShape} at $\sim6.0\%$, \texttt{Translient} at $\sim5.8\%$ and \texttt{Scorr} at $\sim6.1\%$. These rejection rates are dominated by lower-flux injections that are harder to distinguish from the contribution of their galaxy hosts in the case of \texttt{Scorr}, potential contamination of nearby smeared sources in the case of \texttt{PSFShape}, or statistically degenerate in the case of \texttt{Translient}. The rejection rates decay to $\sim3.7\%$ for \texttt{PSFShape}, $\sim3.8\%$ for \texttt{Translient}, and $\sim2.1\%$ for \texttt{Scorr} when sub-selecting candidates with $S \ge 7.5\sigma$. Host contribution generally increases the rejection rate for faint injections. To estimate the efficiency for non-blended transients, we consider a subset of injections that are at least $6$`` away from any external cataloged source. In this case, the overall survival rate increases to $\sim81$\%. The efficiency curve for this sub-selection is also shown in Figure~\ref{fig:InjectionsEfficiency}. Logistic fits to the efficiency curves yield asymptotic high-significance efficiencies of $82.4^{+0.7}_{-0.8}\%$, $87.0^{+0.6}_{-0.8}\%$, and $86.2^{+1.0}_{-1.0}\%$ for the full, no-crop, and non-blended samples, respectively, where the uncertainties are derived from bootstrap resampling.

We emphasize that the single-epoch efficiency should be interpreted as a lower bound on the probability of detecting a real transient. Since astrophysical transients are typically observed over multiple epochs, each event has several independent chances to pass the filtering stage. The cumulative detection probability therefore increases with the number of epochs, partially compensating for single-epoch inefficiencies. This is particularly important for high-cadence surveys such as LAST, where even moderate single-epoch efficiencies can translate into high overall recovery rates.

\subsection{Runtime summary.}
\label{sec:runtime}

Pipeline~II runtime scales primarily with the number of detected initial candidates $N_c$, approximately linearly with a mild superlinear component. A simple linear fit gives an intercept of $\sim45\,$s and a per-candidate cost of $\sim1.24$\,ms ($R^2\simeq0.93$), while allowing a weak quadratic or power-law term improves the fit slightly ($R^2\simeq0.94$) and suggests $t\propto N_c^{1.2}$. For a typical nightly load of $\sim1.7\times10^6$ candidates per telescope and $6$\,h on sky, the estimated total runtime per telescope is $\sim1.2$\,h, that is, under $20\%$ of the night. Full methodology, fitted coefficients, and benchmarks are provided in Appendix~\ref{app:runtime}.

\section{Conclusion}
\label{sec:conclusion}

We presented the second part of the LAST pipeline (Pipeline~II), which performs image subtraction and transient detection on the LAST data stream in real time. Pipeline~II operates on the calibrated $20\times20\,$s coadded images produced by the first part of the pipeline (Pipeline~I) and carries out reference image registration, subtraction based on the ZOGY algorithm, and detection of transient sources based on the subtraction statistic $S$. The subtraction process is augmented by the \textsc{Translient} statistic, which is employed to distinguish true flux variations from small astrometric shifts, a Gabor-filter based method for identifying Fourier ringing artifacts, and a source-noise correction $S_\text{corr}$ for determining overestimation of $S$ on bright sources.  

Each transient candidate with $\vert S\vert\ge5\sigma$ undergoes aperture photometry and PSF photometry in the reference, new, and difference images, as well as a series of morphological and statistical quality tests. Additional filters are employed to identify and remove artifacts such as bad pixels, streaks, ringing, and poor PSF reconstructions, while cross-matching eliminates known variables, stars, and Solar System objects. The pipeline is fully deterministic, producing a clean, high-purity stream of candidates suitable for classification and follow-up.

From the commissioning data, we have found a $5\sigma$ single-epoch limiting magnitude of about $20.3$\,mag for a standard $20\times20\,$s reference and about $20.7$\,mag with deeper references. As measured in a sample of $\approx4\times10^{7}$ initial detections, the filtering pipeline reduces the number of candidates down to a transients density of $\sim0.01$\,deg$^{-2}$ with a purity of $\gtrsim90\%$ at $S\geq7.5\sigma$. The single-epoch efficiency reaches $\sim80\%$ in the full sample and up to $\sim87\%$ under favorable conditions. The dominant systematic sources of remaining inefficiency are registration trims and PSF reconstruction limitations. Future improvements will include refined reference registration, enhanced PSF modeling, and potential incorporation of machine-learning classifiers which will operate on the parameter space defined so far.

\section*{Acknowledgments}

R.K. is grateful for receiving the Dean of Faculty fellowship. E.O.O. is grateful for the support of grants from the Willner Family Leadership Institute, André Deloro Institute, Paul and Tina Gardner, The Norman E Alexander Family, M Foundation, ULTRASAT Data Center Fund, Israel Science Foundation, Israeli Ministry of Science, Minerva, BSF, BSF-transformative, NSF-BSF, Israel Council for Higher Education (VATAT), Sagol Weizmann-MIT, Yeda-Sela, and Weizmann-UK. S.A.S. is grateful for the Zuckerman Scholars fellowship. 

\bibliography{papers2.bib}
\bibliographystyle{aa}

\appendix
\nolinenumbers
\section{Pipeline~I overview}
\label{sec:appendix_pipeI}

The following summarizes the main steps of Pipeline~I (see also \citealt{Ofek+2023PASP_LAST_PipeplineI}):

\begin{enumerate}
    \item Create a 32-bit-mask image for the science image, and flag saturated pixels in the mask image (see Table~1 in \citealt{Ofek+2023PASP_LAST_PipeplineI}).
    \item Subtract a master dark image from the raw science image. The dark image has its own bit-mask image, which is propagated using the {\it or} operator to the bit-mask image of the science image.
    \item Subtract the overscan dark value from the image.
    \item Divide the science image by the most recent flat field image and propagate the bit mask of the flat field image into the science image bit mask.
    \item Trim the overscan area in the camera.
    \item Multiply the pixel values by a constant gain.
    \item Partition the image into 24 sub-images. The sub-images have a size of $1726\times 1726$ pixels,
    including an at least 64-pixel-wide overlap between sub-images. The overlap is important to avoid losing sources and to overcome pointing inaccuracies. From this step forward, all the processing is done independently on individual sub-images.
    \item Measure the background and variance images for each sub-image.
    \item Subtract the background from each sub-image.
    \item Match filter the image using two Gaussian filters with sigma-widths of 0.1 and 1.5 pixels.
    \item Find all sources with a signal-to-noise ratio $S/N \geq 5$ in each match-filtered image.
    \item Compare the $S/N$ corresponding to the delta function (the 0.1 pixel filter) with that of wider filters for each source. Remove all sources for which the delta function filter scores a higher $S/N$. The positions of removed sources are marked as cosmic ray hits in the bit mask. 
    \item Perform aperture photometry and retrieve the source flux as well as the background and its standard deviation in an annulus around each source. Additionally, derive the first and second moments of the source PSF for each source.
    \item Propagate the bit-mask information into the source catalog. This includes the overlap and near-edge bits. Specifically, the overlap bit allows us to remove duplicate sources found in the overlap region.
    \item Select sources with $S/N \in [50, 1000]$, second moments in the $[10, 80]$ percentile, and no neighbors within six pixels. These sources are shifted using cubic interpolation to the center of the PSF stamp and coded to produce the sub-image PSF.
    \item Perform PSF fit photometry for all sources.
    \item Solve the astrometry for each of the sub-image fields and update the corresponding source catalogs with the astrometric information.
    \item Perform photometric calibration of the sources against GAIA-DR3 sources (\citealt{GAIA+2016_GAIA_mission, GAIA+2022yCat_GAIA_DR3_MainSourcesCatalog}). The calibration includes a zero point and a color term and achieves a typical accuracy of about 0.015\,mag. The calibrated magnitudes are evaluated assuming a color of $B_{\rm p}-R_{\rm p}=1$\,mag. Additionally, determine the image zero point and limiting magnitude for a 5$\,\sigma$ detection.
    \item Match the sources in the 20 successive sub-images of a visit and produce a matched source catalog that contains all detected sources.
    \item Fit each matched source with a proper motion model and check for changes in flux.
    \item Search for asteroids in the 20 successive images.
    \item Match the matched sources with known solar system objects.
    \item Match the matched sources with external catalogs by using catsHTM \citep{Soumagnac+Ofek2018_catsHTM}.
    \item Transform the 20 successive sub-images into the same frame and coadd them.
    \item Produce a calibrated source catalog for each of the 24 coadded sub-images using the same method as individual epoch sub-images (steps 8--16).
    \item Refine the astrometry of the coadded source catalog using the PSF-fit positions.
    \item Save data products to disk. The data products include images, catalogs, mask images, and PSF stamps (see Table~7 in \citealt{Ofek+2023PASP_LAST_PipeplineI}).
    \item Save all produced source catalogs and headers of the coadded images to a database.
\end{enumerate}

In the near future, steps 10 to 16 will be replaced with a multi-iteration matched filtering and PSF subtraction including the PSF, a delta function, and an extended PSF. Initial implementations produce more than twice the number of detected sources in dense regions, such as the Galactic Plane. Step 18 will replaced by photometric calibration based on the system-transmission as described in \cite{SimonePhotometryCalib}.

\section{Candidate reduction by single filters}
\label{app:singlefilters}

\begin{figure}
\includegraphics[width=0.49\textwidth]{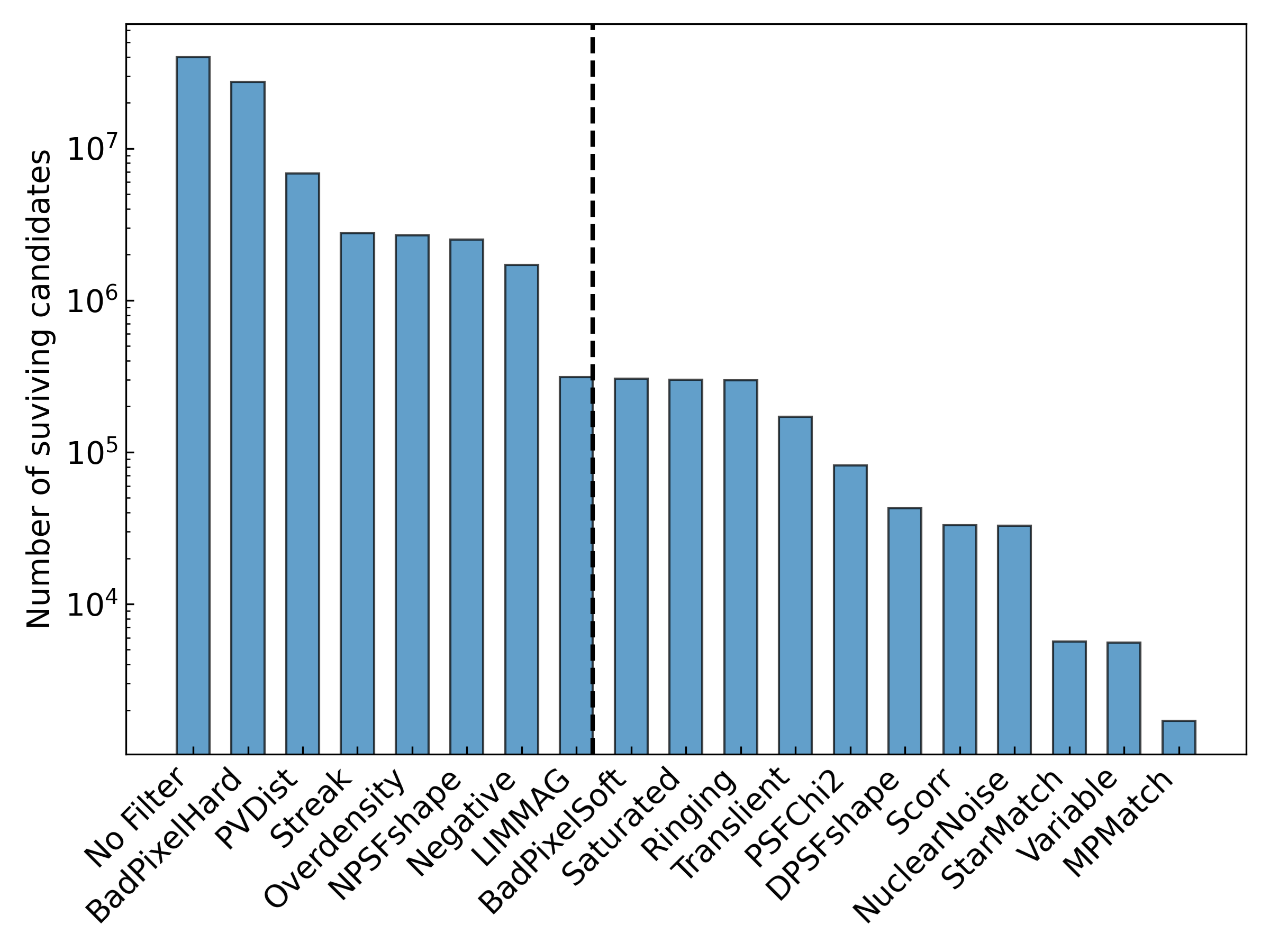}%
\hfill \\
\includegraphics[width=0.49\textwidth]{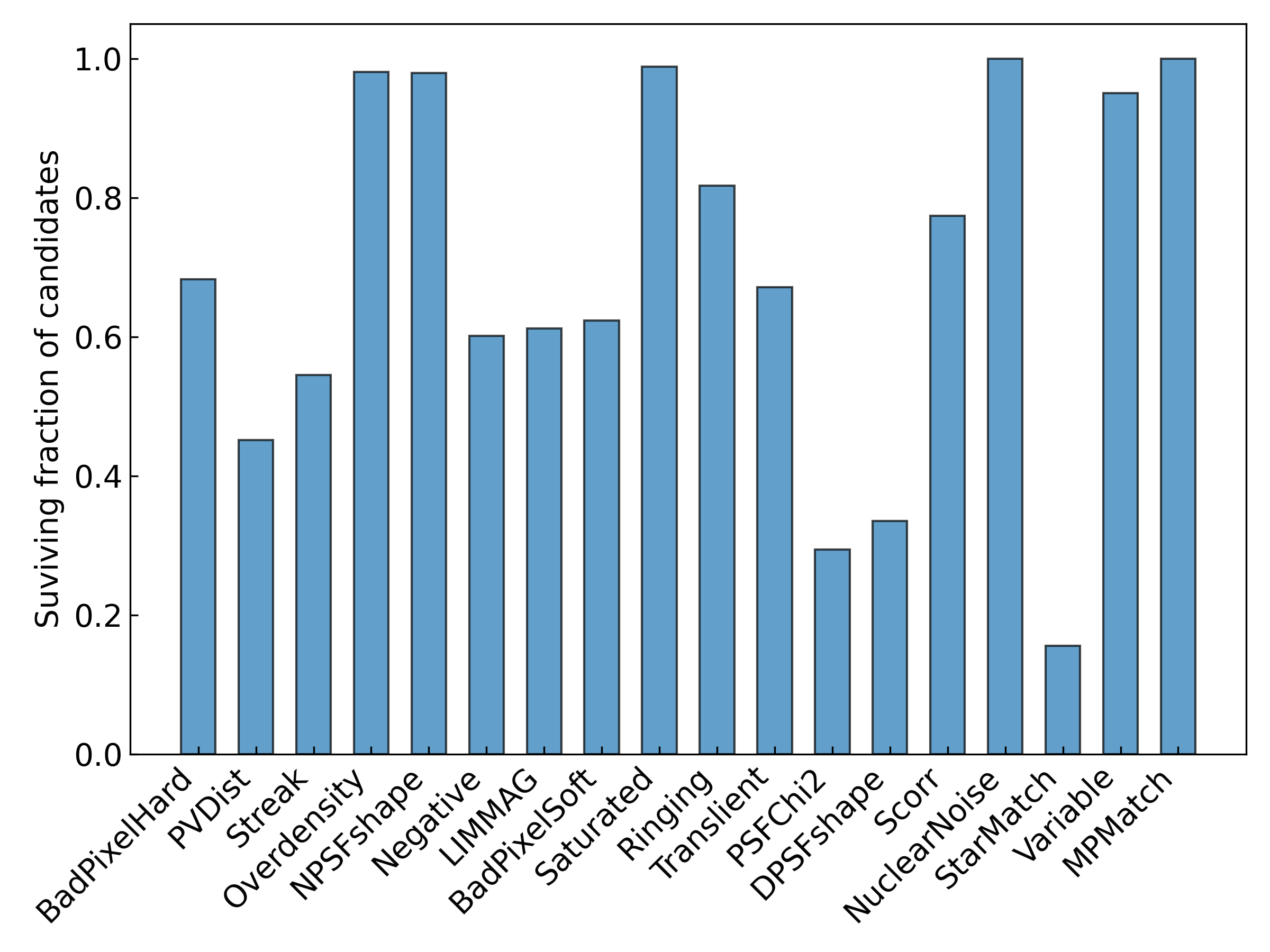}
\caption{Filtering performance of the single-epoch filters. The top panel shows the surviving number of candidates after sequential application of all filters. The vertical dashed line marks the level~2 sub-selection, after which candidates are not retained. The bottom panel shows the surviving fraction when each filter is applied independently.}
\label{fig:SequantialAndIndividualFilters}
\end{figure}

\begin{figure}
\centering
\includegraphics[width=0.48\textwidth]{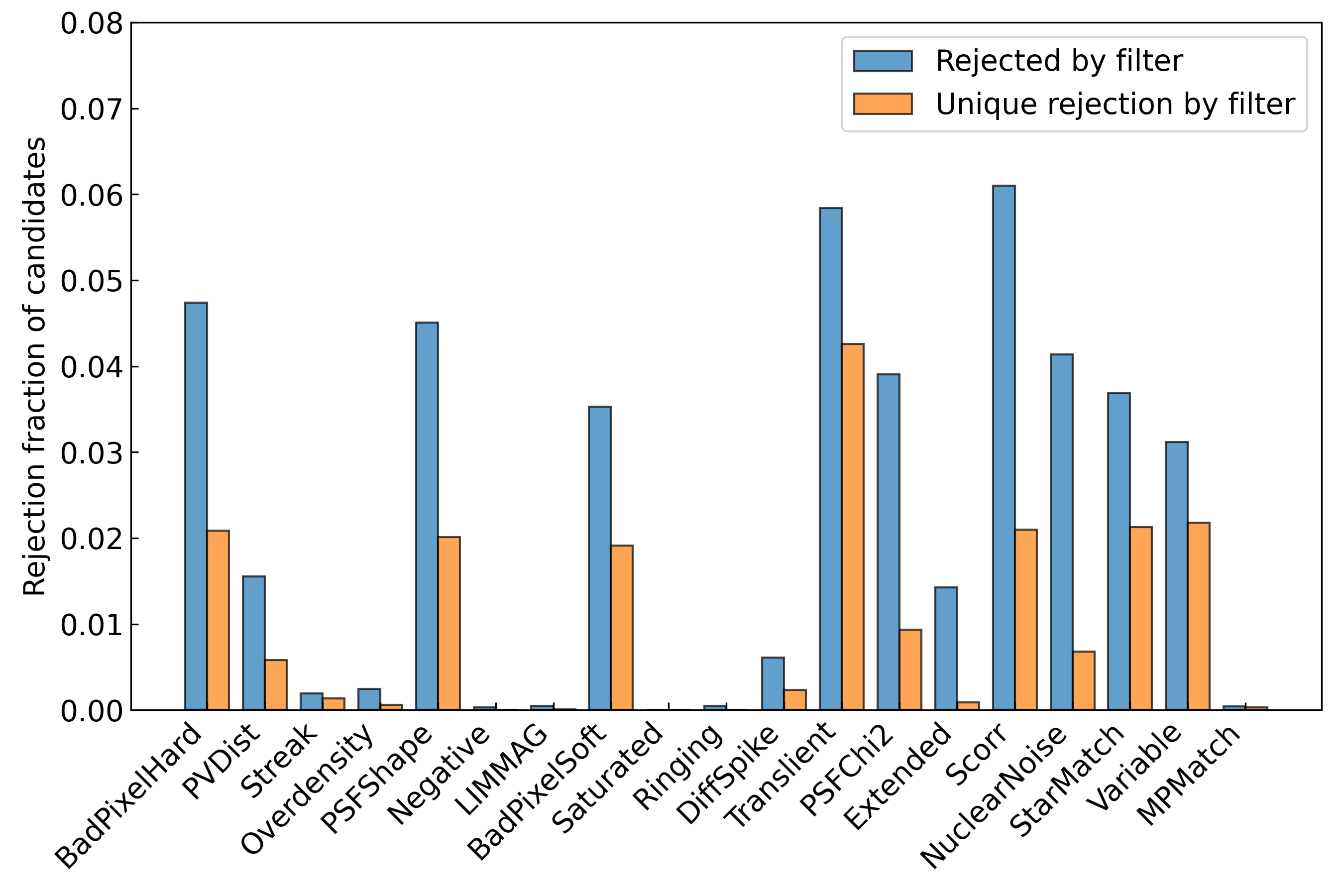}
\caption{Rejection fraction of single-epoch filters applied to artificially injected transients on top of galaxies.}
\label{fig:InjectionsIndividualFilters}
\end{figure}

Figure \ref{fig:SequantialAndIndividualFilters} (top) shows the sequential application of all single-epoch filters, where each bar represents the number of candidates surviving after each filter step. For clarity, the filters are shown ordered from those addressing artifacts to those targeting astrophysical contaminants. The sub-selection of level 2 candidates is shown by a dashed vertical line, i.e., all candidates that fail up to the marked point are not saved within the database. The level 2 sub-selection thus decreases the number of candidates by two orders of magnitude. In this case, the number of candidates is reduced from $\sim40\,$million to $\sim300\,000$, a retention rate of $\sim0.75\%$. Figure \ref{fig:SequantialAndIndividualFilters} (bottom), in contrast, presents the effect of applying each filter independently, showing the surviving fraction of the initial candidate set. The filtering rates indicate that the dominant sources of false positive candidates are: (1) persistent sources; (2) poor PSFs or PSF reconstructions; (3) single-pixel peaks—such as those caused by cosmic rays, bad pixels, and statistical fluctuations; and (4) registration errors. Figure~\ref{fig:InjectionsIndividualFilters} shows the rejection fraction for individual filters applied only artificially injection transients.

\section{Runtime}
\label{app:runtime}
\begin{figure}
    \centerline{\includegraphics[width=0.48\textwidth]{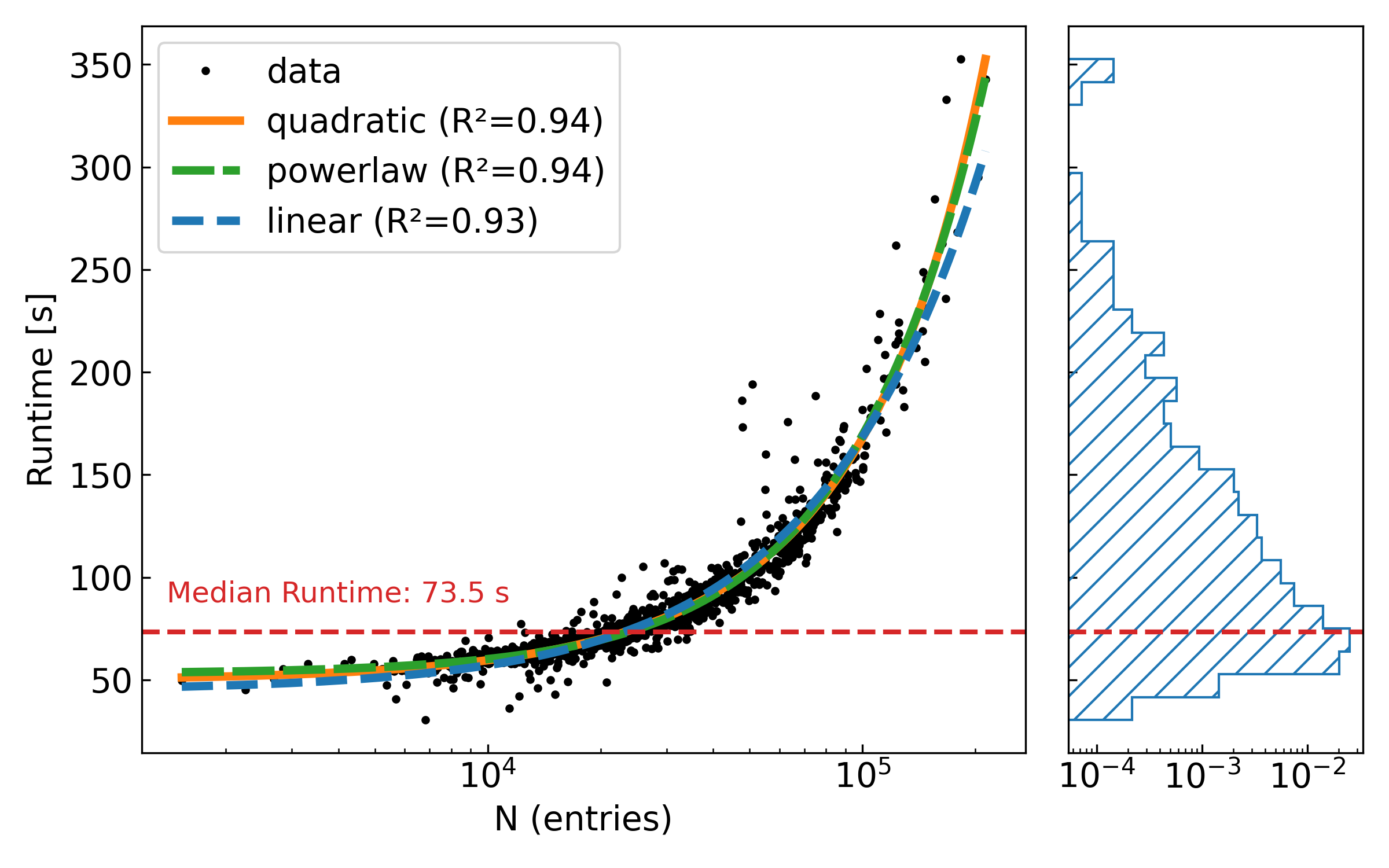}}
    \caption{The runtime of Pipeline~II measured for a sample of 1244 invocations. The left panel shows the runtime as a function of the number of detected initial candidates $N_c$. The data are shown as black points, while a quadratic fit is shown as a solid orange line, a power-law fit is shown as a green dashed line, and a linear fit is shown as a sparsely dashed blue line. The fit results are listed in Table~\ref{tab:runtimefits}. The right panel shows the density histogram of the measured runtime. The median runtime is shown as a horizontal dashed red line.}
    \label{fig:Runtime}
\end{figure}

The runtime of Pipeline~II depends primarily on the number of candidates processed per invocation. Figure~\ref{fig:Runtime} shows the measured wall time as a function of the detected initial candidates $N_c$, along with linear, quadratic, and power-law fits (parameters in Table~\ref{tab:runtimefits}). The timing sample uses 1244 single-telescope observations. To emulate real-time performance, all $N$ images were preloaded prior to each call. The reported time includes processing 24 calibrated LAST sub-images of size $1726\times1726$ pix and writing all products to disk, but excludes post-processing steps such as cleaning and database injection as shown in Figure~\ref{fig:LAST_transients_pipe}. All benchmarks were run on a workstation with an AMD EPYC~7763 CPU (64 cores, 128~threads), 503~GiB RAM, and NVMe storage, running Ubuntu~22.04.5 LTS (kernel 6.8.0–85–generic). No parallelization was employed beyond MATLAB internal processes.

The measured runtimes scale approximately linearly with $N_c$, with a weak superlinear component. A linear model yields an intercept of $(44.77\pm0.42)\,$s and a per-candidate cost of $\sim1.24\,$ms ($R^2\simeq0.93$). Allowing for a quadratic or power-law term improves the fit slightly to $R^2\simeq0.94$, with a best-fit exponent of $1.213\pm0.017$ and a small quadratic coefficient $(2.15\pm0.17)\times10^{-9}$, implying $t\propto N_c^{1.2}$. The $\approx50\,$s constant overhead reflects initialization and I/O, dominated by generating the $D$, $S$, $S_\text{corr}$, \textsc{Translient}, and related images and by registering the $N$ and $R$ frames. The weak nonlinearity becomes relevant only at high densities, $N_c\gtrsim10^5$.

For a $6\,$h night with up to 54 visits per telescope, the nightly mean of $\sim1.7\times10^6$ candidates implies total Pipeline~II runtimes of $(1.255\pm0.007)\,$h (linear), $(1.22\pm0.01)\,$h (quadratic), and $(1.2\pm0.1)\,$h (power law). Thus, real-time processing occupies $\lesssim20\%$ of the nightly budget. Since each telescope is processed independently, these estimates do not depend on the total number of telescopes. Lower runtimes are foreseen with further development.

\begin{table*}
\caption{Runtime fit parameters for Pipeline~II as a function of candidate count\label{tab:runtimefits}}
\centering
\begin{tabular}{lcccccc}
\hline\hline
Model &
Intercept [s] &
$b$ &
$c$ or $p$ &
RMSE [s] &
$R^2$ &
AIC / BIC \\
\hline
Linear    & $44.77\pm 0.42$ & $(1.236\pm 0.010)\times10^{-3}$ & - & $9.46$ & $0.930$ & $5593.5\,/\,5603.8$ \\
Quadratic & $49.66\pm 0.55$ & $(9.67\pm0.23)\times10^{-4}$ & $(2.15\pm 0.17)\times10^{-9}$ & $8.90$ & $0.938$ & $5444.6\,/\,5460.0$ \\
Power law & $53.01\pm 0.66$ & $(9.98\pm 2.01)\times10^{-5}$ & $1.213\pm 0.017$ & $8.91$ & $0.938$ & $5447.8\,/\,5463.2$ \\
\hline
\end{tabular}
\tablefoot{\parbox{0.95\textwidth}{%
    \textit{Columns:} (1): functional form used to fit the runtime as a function of the candidate count $N_c$. (2): fitted constant offset, representing pipeline initialization, I/O overhead, and fixed set-up. (3) and (4): model parameters with $1\sigma$ uncertainties. For the linear and quadratic fits, $b$ and $c$ correspond to the coefficients of the $N_c$ and $N_c^2$ terms, respectively, while for the power-law fit, $p$ is the fitted exponent in $t=a+bN_c^p$. (5) and (6): root mean squared error (RMSE) of the residuals and coefficient of determination ($R^2$), describing the goodness of fit. (7): Akaike Information Criterion (AIC) and Bayesian Information Criterion (BIC), which balance model complexity and fit quality. Lower values indicate a better trade-off. 
    The quadratic model yields the lowest AIC and BIC values, indicating a weak but significant superlinear scaling component.
}}
\end{table*}

\FloatBarrier
\section{Level 2 transients catalog columns}
\label{app:fullcatalog}

Table~\ref{tab:full} lists all saved parameter columns per candidate at level $2$ as saved within the transients database.

\longtab[1]{%
\setlength{\tabcolsep}{6pt}%
\begin{longtable}{lcl}
\caption{Level 2 transients catalog columns.\label{tab:full}}\\
\hline\hline
Column name & Units & Description \\
\hline
\endfirsthead
\caption{Level 2 transients catalog columns (continued).}\\
\hline\hline
Column name & Units & Description \\
\hline
\endhead
\hline
\endfoot
\hline
\multicolumn{3}{l}{$^*$ corrected Akaike information criterion.}\\
\endlastfoot
\texttt{id\_diff\_src} &  & Difference image source ID \\
\texttt{id\_uniq\_src} &  & Source ID in the unique sources table \\
\texttt{id\_cat\_src}  &  & Source ID in the sources table of the new image\\
\texttt{id\_ref\_im}   &  & $R$ image identifier \\
\texttt{id\_new\_im}   &  & $N$ image identifier \\
\texttt{upix\_partition} &  & Uniq healpix ID of Nside $2^3$ \\
\texttt{upix\_low}     &  & Uniq healpix ID of Nside $2^8$ \\
\texttt{upix\_high}    &  & Uniq healpix ID of Nside $2^{16}$ \\
\texttt{is\_forced}    &  & The entry is based on a forced photometry request \\
\texttt{score}         & $\sigma$ & ZOGY $S$ significance \\
\texttt{s\_corr}       & $\sigma$ & Source noise corrected $S$ \\
\texttt{sn\_delta}     & $\sigma$ & S/N with a $\delta$-function kernel \\
\texttt{s2}            &  & Auxiliary statistic S2 \\
\texttt{s2\_sig}       & $\sigma$ & Significance of S2 \\
\texttt{s2\_aic}       &  & AIC$_\text{c}^*$ for the S2 model \\
\texttt{z2}            &  & \texttt{Translient} statistic \\
\texttt{z2\_sig}       & $\sigma$ & Significance of \texttt{Translient} \\
\texttt{z2\_aic}       &  & AIC$_\text{c}^*$ for the \texttt{Translient} \\
\texttt{n\_sn}         &  & Forced photometry S/N in $N$ \\
\texttt{n\_psf\_chi2dof} &  & PSF fit $\chi^{2}$/dof in $N$ \\
\texttt{n\_flux\_psf}  & e$^{-}$ & PSF flux in $N$ \\
\texttt{n\_fluxerr\_psf} & e$^{-}$ & Uncertainty of the PSF flux in $N$ \\
\texttt{n\_mag\_psf}     & mag & PSF magnitude in $N$ \\
\texttt{n\_magerr\_psf}  & mag & Uncertainty of the PSF magnitude in $N$ \\
\texttt{r\_sn}         &  & Forced photometry S/N in $R$ \\
\texttt{r\_psf\_chi2dof} &  & PSF fit $\chi^{2}$/dof in $R$ \\
\texttt{r\_flux\_psf}  & e$^{-}$ & PSF flux in $R$ \\
\texttt{r\_fluxerr\_psf} & e$^{-}$ & Uncertainty of the PSF flux in $R$ \\
\texttt{r\_mag\_psf}     & mag & PSF magnitude in $R$ \\
\texttt{r\_magerr\_psf}  & mag & Uncertainty of the PSF magnitude in $R$ \\
\texttt{peak\_dist}    & pix & Distance from $D$ peak to nearest $N$ peak \\
\texttt{pv\_dist}      & pix & Peak–valley separation $d_{\rm PV}$ \\
\texttt{n\_flags}      &  & Bitmask from $N$ around the candidate \\
\texttt{r\_flags}      &  & Bitmask from $R$ around the candidate \\
\texttt{flags}         &  & Bitmask from $D$ around the candidate \\
\texttt{sn\_gabor}     & $\sigma$ & Gabor score at the candidate location \\
\texttt{flags\_transient} &  & Single epoch flagging bitmask \\
\texttt{r\_jd}         & d   & Mid time of $R$ (JD) \\
\texttt{n\_fwhm}       & pix & PSF FWHM in $N$ \\
\texttt{r\_fwhm}       & pix & PSF FWHM in $R$ \\
\texttt{n\_limmag}     & mag & $5\sigma$ limiting mag in $N$ \\
\texttt{r\_limmag}     & mag & $5\sigma$ limiting mag in $R$ \\
\texttt{n\_zp}         & mag & Zero point of $N$ \\
\texttt{r\_zp}         & mag & Zero point of $R$ \\
\texttt{n\_ph\_col1}   & mag & Color term used for $N$ calibration \\
\texttt{r\_ph\_col1}   & mag & Color term used for $R$ calibration \\
\texttt{n\_exptime}    & s   & Exposure time of $N$ \\
\texttt{r\_exptime}    & s   & Exposure time of $R$ \\
\texttt{n\_distmp}     & arcsec & Distance to the nearest minor planet at the epoch of $N$ \\
\texttt{n\_magmp}      & mag & Predicted mag of the nearest minor planet at the epoch of $N$\\
\texttt{r\_distmp}     & arcsec & Distance to the nearest minor planet at the epoch of $R$ \\
\texttt{r\_magmp}      & mag & Predicted mag of the nearest minor planet at the epoch of $R$\\
\texttt{jd}            & d  & Detection time (JD), the mid time of $N$ \\
\texttt{bjd}           & d  & Barycentric JD \\
\texttt{baryvel}       & km s$^{-1}$ & Barycentric velocity used \\
\texttt{ra}            & deg & Right ascension (ICRS, J2000) \\
\texttt{dec}           & deg & Declination (ICRS, J2000) \\
\texttt{xpeak}         & pix & X of the maximum in $S$ (pixel precision) \\
\texttt{ypeak}         & pix & Y of the maximum in $S$ (pixel precision) \\
\texttt{x1}            & pix & X of the maximum in $S$ (subpixel precision) \\
\texttt{y1}            & pix & Y of the maximum in $S$ (subpixel precision) \\
\texttt{n\_x2}         & pix$^{2}$ & $\langle x_{N}^{2}\rangle$ of $P_N$ \\
\texttt{n\_y2}         & pix$^{2}$ & $\langle y_{N}^{2}\rangle$ of $P_N$ \\
\texttt{n\_xy}         & pix$^{2}$ & $\langle x_{N}*y_{N}\rangle$ of $P_N$ \\
\texttt{r\_x2}         & pix$^{2}$ & $\langle x_{R}^{2}\rangle$ of $P_R$ \\
\texttt{r\_y2}         & pix$^{2}$ & $\langle y_{R}^{2}\rangle$ of $P_R$ \\
\texttt{r\_xy}         & pix$^{2}$ & $\langle x_{R}*y_{R}\rangle$ of $P_R$ \\
\texttt{flux\_aper\_1} & e$^{-}$ & Aperture flux in $D$, r=2 pix \\
\texttt{fluxerr\_aper\_1} & e$^{-}$ & Uncertainty of the aperture flux in $D$, r=2 pix \\
\texttt{mag\_aper\_1}  & mag & Aperture mag in $D$, r=2 pix \\
\texttt{magerr\_aper\_1} & mag & Uncertainty of the aperture mag in $D$, r=2 pix \\
\texttt{flux\_aper\_2} & e$^{-}$ & Aperture flux in $D$, r=4 pix \\
\texttt{fluxerr\_aper\_2} & e$^{-}$ & Uncertainty of the aperture flux in $D$, r=4 pix \\
\texttt{mag\_aper\_2}  & mag & Aperture mag in $D$, r=4 pix \\
\texttt{magerr\_aper\_2} & mag & Uncertainty of the aperture mag in $D$, r=4 pix \\
\texttt{flux\_aper\_3} & e$^{-}$ & Aperture flux in $D$, r=6 pix \\
\texttt{fluxerr\_aper\_3} & e$^{-}$ & Uncertainty of the aperture flux in $D$, r=6 pix \\
\texttt{mag\_aper\_3}  & mag & Aperture mag in $D$, r=6 pix \\
\texttt{magerr\_aper\_3} & mag & Uncertainty of the aperture mag in $D$, r=6 pix \\
\texttt{sn}            &  & Forced photometry S/N in $D$ \\
\texttt{flux\_psf}     & e$^{-}$ & PSF flux in $D$ \\
\texttt{fluxerr\_psf}  & e$^{-}$ & Uncertainty of the PSF flux in $D$ \\
\texttt{mag\_psf}      & mag & PSF magnitude in $D$ \\
\texttt{magerr\_psf}   & mag & Uncertainty of the PSF magnitude in $D$ \\
\texttt{back\_im}      & e$^{-}$ pix$^{-1}$ & Local background in $D$ \\
\texttt{var\_im}       & e$^{-2}$ pix$^{-1}$ & Local variance in $D$ \\
\texttt{back\_annulus} & e$^{-}$ pix$^{-1}$ & Annulus background in $D$ \\
\texttt{std\_annulus}  & e$^{-}$ pix$^{-1}$ & Annulus background standard deviation in $D$ \\
\texttt{gal\_n}        &  & Number of matched galaxies \\
\texttt{gal\_dist}     & arcsec & Distance to the nearest matched galaxy \\
\texttt{star\_n}       &  & Number of matched stars \\
\texttt{star\_dist}    & arcsec & Distance to the nearest matched star \\
\texttt{object}        &  & Observed object of $N$ image, e.g. LAST field ID \\
\texttt{nodenumb}      &  & Node identifier \\
\texttt{mountnum}      &  & Mount identifier \\
\texttt{camnum}        &  & Camera identifier \\
\texttt{cropid}        &  & Sub-image index within the full frame \\
\texttt{zp}            & mag & Zero point used for $D$ photometry \\
\texttt{report\_jd}    & d  & Time stamp of internal report (JD) \\
\texttt{x2}            & pix$^{2}$ & $\langle x^{2}\rangle$ of residual in $D$ \\
\texttt{y2}            & pix$^{2}$ & $\langle y^{2}\rangle$ of residual in $D$ \\
\texttt{xy}            & pix$^{2}$ & $\langle xy\rangle$ of residual in $D$ \\
\texttt{ingestion\_time\_jd} & d & Time stamp of database injection (JD) \\
\texttt{limmag}        & mag & $5\sigma$ limiting mag of $D$ \\
\texttt{n\_neigh}      &  & Number of neighbors \\
\texttt{density}       & pix$^{-1}$ & Local neighbor density \\
\texttt{r\_psf\_chi2dof\_med} &  & Median PSF $\tilde{\chi}^{2}$/dof vs mag in $R$ \\
\texttt{n\_psf\_chi2dof\_med} &  & Median PSF $\tilde{\chi}^{2}$/dof vs mag in $N$ \\
\texttt{dsdf}          &  & Derivative of $S$ w.r.t. flux ratio \\
\texttt{flux\_contam}  & e$^{-}$ & Estimated contaminating flux from neighbors \\
\texttt{gaia\_bp}      & mag & Gaia $B_{\rm p}$ magnitude if matched \\
\texttt{gaia\_rp}      & mag & Gaia $R_{\rm p}$ magnitude if matched \\
\end{longtable}
}

\end{document}